\documentclass{article}

\usepackage{arxiv}
\usepackage[utf8]{inputenc}
\usepackage[T1]{fontenc}
\usepackage{hyperref}
\usepackage{url}
\usepackage{booktabs}
\usepackage{amsmath,amssymb,amsfonts}
\usepackage{graphicx}
\usepackage{microtype}
\usepackage[numbers,sort&compress]{natbib}
\usepackage{cleveref}
\usepackage{enumitem}
\usepackage{optidef}
\usepackage[acronym]{glossaries}
\usepackage{siunitx}
\usepackage{authblk}
\usepackage[version=4]{mhchem}
\usepackage{longtable,tabularx}
\usepackage{adjustbox}
\usepackage{algorithm}
\usepackage[noend]{algpseudocode}

\setlist[itemize]{leftmargin=1.5em}
\setlist[enumerate]{leftmargin=1.7em}
\title{Station-Keeping Approach for Extremely Low Lunar Orbits with Solar Sailing}
\date{}
\glsdisablehyper

\author[1]{%
	{Jack Yarndley\thanks{Corresponding Author: \texttt{jyar540@aucklanduni.ac.nz}}}%
}
\author[2]{%
	{Gregory Lantoine}%
}
\author[1]{%
	{Roberto Armellin}%
}

\affil[1]{Te P\=unaha \=Atea -- Space Institute, University of Auckland, Auckland 1010, New Zealand}
\affil[2]{Jet Propulsion Laboratory, California Institute of Technology, Pasadena, CA 91109, USA}

\renewcommand{\headeright}{}
\renewcommand{\undertitle}{}
\renewcommand{\shorttitle}{\textit{Station-Keeping Approach for Extremely Low Lunar Orbits with Solar Sailing}}

\makeatletter
\renewcommand\maketitle{%
    {{%
        \renewenvironment{tabular}[2][]{%
            \begin{center}\begin{minipage}{0.8\textwidth}\centering
        }{%
            \end{minipage}\end{center}
        }%
        \AB@maketitle
    }}%
}
\makeatother

\hypersetup{
pdftitle={Station-Keeping Approach for Extremely Low Lunar Orbits with Solar Sailing},
pdfsubject={physics.space-ph},
pdfauthor={Jack Yarndley, Gregory Lantoine and Roberto Armellin},
pdfkeywords={solar sailing, extremely low lunar orbits, station-keeping, sequential convex programming, lossless convexification, eccentricity vector},
}

\newacronym{SCVX}{SCVX}{successive convex optimization}
\newacronym{SCP}{SCP}{sequential convex programming}
\newacronym{FOH}{FOH}{first-order-hold discretization}
\newacronym{LLO}{LLO}{low-altitude lunar orbit}
\newacronym{eLLO}{eLLO}{extremely low-lunar orbit}
\newacronym{ZOH}{ZOH}{zero-order-hold}
\newacronym{LRO}{LRO}{Lunar Reconnaissance Orbiter}
\newacronym{ICRF}{ICRF}{international celestial reference frame}
\newacronym{DCM}{DCM}{direction cosine matrix}
\newacronym{LME2000}{LME2000}{lunar mean equator of date J2000}
\newacronym{GRAIL}{GRAIL}{Gravity Recovery And Interior Laboratory}
\newacronym{SDF}{SDF}{Signed Distance Field}
\newacronym{AD}{AD}{automatic differentiation}
\newacronym{SRP}{SRP}{solar radiation pressure}
\newacronym{DDP}{DDP}{differential dynamic programming}
\newacronym{MEE}{MEE}{modified equinoctial element}
\newacronym{SOC}{SOC}{second-order cone}
\newacronym{MIQCP}{MIQCP}{mixed-integer quadratically constrained programming}
\newacronym{MISOCP}{MISOCP}{mixed-integer second-order cone programming}

\begin{document}
\maketitle

\begin{abstract}
Renewed interest in cislunar space has created opportunities for sustained operations in extremely low-lunar orbits (eLLOs), where altitudes below 50~km enable close surface proximity. However, these orbits are strongly perturbed by the irregular lunar gravity field, leading to rapid eccentricity growth, high station-keeping costs or even surface impact. Recent advances in our understanding of the lunar `translation theorem' have revealed predictable behavior in the eccentricity vector, offering new opportunities for efficient control. This paper introduces a two-stage framework for solar sail station-keeping in eLLOs. First, a mixed-integer second-order cone programming (MISOCP) approach leverages the translational behavior of the eccentricity vector to identify orbit and sail configurations favorable for station-keeping. Second, a lightweight sequential convex programming (SCP) formulation refines these into high-fidelity trajectories, enabled by a recently developed lossless convexification of solar sail dynamics. A case study inspired by the Lunar Reconnaissance Orbiter (LRO) mission demonstrates that a realistic solar sail spacecraft can be maintained within the eLLO regime for at least 1~year without propellant expenditure, suggesting that longer-duration, or even indefinite station-keeping, may be feasible. The approach remains effective at reduced control update frequencies (down to monthly) and exhibits low sensitivity to uncertainties.
\end{abstract}

\glsresetall

\begin{center}
\textbf{Keywords:} solar sailing, extremely low lunar orbits, station-keeping, sequential convex programming, lossless convexification, eccentricity vector
\end{center}

\section{Introduction}
\label{sec_introduction}

Cislunar space has attracted renewed interest in recent years, driven by an increasing number of governmental and commercial missions. One particularly attractive region of cislunar space is \glspl{eLLO}, with average lunar orbital altitudes below 50~km. Unlike orbits around Earth, these orbits are viable due to the absence of a lunar atmosphere and therefore present unique scientific and technological opportunities. However, the orbit and station-keeping design of \glspl{eLLO} is particularly challenged by the highly non-spherical lunar gravity field, an effect that becomes more pronounced at lower altitudes. This gravity field induces rapid eccentricity growth in \glspl{eLLO}, which, if left unchecked, will lead to surface impact. Therefore, careful design of orbital configuration and station-keeping strategies is required \citep{beckmanStationkeepingLunarReconnaissance2007, mesarchManeuverOperationsResults2010, wallaceLowLunarOrbit2012, yarndleyOriginsApplicationTranslation2026}, and even then, the associated station-keeping costs are generally high.

The lunar gravitational field is often represented using a spherical harmonic expansion, such as in models from the \gls{GRAIL} mission \citep{konoplivJPLLunarGravity2013}. At high order and degree, these models provide excellent fidelity but are computationally expensive to evaluate. Therefore, for orbit analysis over longer timescales, these models are generally simplified to retain only terms that drive long-term dynamics. Examples include truncation to zonal harmonic models \citep{hootsHistoryAnalyticalOrbit2012}, classical averaging \citep{brouwerSolutionProblemArtificial1959, kozaiMotionCloseEarth1959, kaulaTheorySatelliteGeodesy2000}, or further simplifications by double averaging over body rotation \citep{metrisSemianalyticalTheoryMean1995, deleflieLongPeriodVariationsEccentricity2006, palacianDynamicsSatelliteOrbiting2007}. Although these simplifications can accurately model long-term orbital dynamics at higher lunar altitudes, in \glspl{eLLO} the removed short-period terms can have significant effects \citep{beckmanStationkeepingLunarReconnaissance2007, yarndleyOriginsApplicationTranslation2026}.

Additionally, due to the slow lunar rotation rate, averaging methods over the body rotation are generally less effective than for most other bodies \citep{kaulaTheorySatelliteGeodesy2000, laraDesignLonglifetimeLunar2011}. This limitation is further compounded by the comparable magnitude of the zonal oblateness coefficient ($J_2$) of the Moon to other sectoral and tesseral coefficients. The result is a significant monthly-recurring modulation in the eccentricity vector that forms recognizable patterns. This behavior was first noted during the \gls{LRO} station-keeping operations \citep{beckmanStationkeepingLunarReconnaissance2007} and referred to as the `translation theorem'. This theorem enables prediction of the eccentricity vector evolution across different initial conditions without numerical reintegration, and has since been dynamically justified by decomposing the motion into mean and monthly components \citep{yarndleyOriginsApplicationTranslation2026}. This relationship provides a useful foundation for developing station-keeping strategies in regions where this effect is pronounced, such as \glspl{eLLO}.

Previous studies on station-keeping in lower lunar orbits have focused on identifying stable frozen or quasi-frozen configurations that mitigate the effects of the lunar gravitational perturbations \citep{nieLunarFrozenOrbits2018}. However, such configurations tend to become impractical or even impossible at altitudes below 50~km \citep{elipeFrozenOrbitsMoon2003, foltaLunarFrozenOrbits2006, russellLongLifetimeLunarRepeat2007, abadAnalyticalModelFind2009, laraDesignLonglifetimeLunar2011, san-juanHighFidelitySemianalyticalTheory2019}. As a result, many studies on station-keeping strategies closer to the lunar surface, such as those for the \gls{LRO}, have predominantly addressed higher average altitudes \citep{beckmanStationkeepingLunarReconnaissance2007, davisEnhancedStationKeepingManeuver2018, cordovaalarconAnalysisLifetimeExtension2019, leonardiLowThrustNonlinearOrbit2024, cinelliLunarOrbitsTelecommunication2024, punoContinuousLowAltitude2026} or have relied on simplifications or propulsion systems that are currently not operationally feasible \citep{singhFeasibilityQuasifrozenNearpolar2020, liuLifetimeExtensionUltra2022}. One of the few exceptions is the \gls{GRAIL} mission, which operated within the \gls{eLLO} regime, particularly during its extended and terminal phases \citep{sweetserDesignExtendedMission2012}. Its station-keeping strategy relied on manual graphical manipulation of the eccentricity vector evolution to maintain a prescribed range \citep{wallaceLowLunarOrbit2012}.

Solar sails harness the momentum of photons from the Sun to generate a continuous acceleration without the need for propellant. Their use is therefore attractive in challenging scientific missions in which propellant mass is at a premium \citep{berthetSpaceSailsAchieving2024}. Only in recent years has solar sail technology matured sufficiently to enable the construction and deployment of solar sails at sizes suitable for spacecraft propulsion, as demonstrated by the broad range of recently proposed and flown missions, such as JAXA's IKAROS (2010) \citep{tsudaAchievementIKAROSJapanese2013}, NASA's NanoSail-D (2008) and NanoSail-D2 (2011) \citep{johnsonNanoSailDSolarSail2011}, LightSail 2 (2019) \citep{spencerLightSail2Solar2021}, NASA's Solar Cruiser (2023) \citep{pezentPreliminaryTrajectoryDesign2021}, the ACS-3 project (2024) \citep{wilkieOverviewNASAAdvanced2021}, and NEA Scout (2024) \citep{lantoineTrajectoryManeuverDesign2024}. From a trajectory design perspective, the limitations of solar sails are profound. Their underlying physics induces a nonlinear coupling between the achievable acceleration directions and magnitudes \citep{mcinnesSolarSailing1999}, and more fundamentally, the maximum acceleration from a solar sail is often orders of magnitude lower than chemical and electric propulsion systems \citep{gongReviewSolarSail2019, spencerSolarSailingTechnology2019}. Moreover, sail attitude must often be highly constrained due to thermal, structural, or power considerations \citep{carusoEffectsAttitudeConstraints2020, oguriSolarSailingPrimer2022, lantoineTrajectoryManeuverDesign2024}. As a result, solar sail spacecraft are highly underactuated dynamical systems, and therefore present significant challenges to trajectory design.

Among a wide range of spacecraft trajectory optimization techniques \citep{chaiReviewOptimizationTechniques2019}, such as those based on direct methods, indirect methods, and \gls{DDP}, there has recently been growing interest in \gls{SCP}. \gls{SCP} is a direct method based on convex programming that iteratively solves a sequence of convex subproblems that approximate the original non-convex problem \citep{maoSuccessiveConvexificationSuperlinearly2019, malyutaConvexOptimizationTrajectory2022}. The main advantages of \gls{SCP} lie in the efficiency, robustness, and convergence guarantees provided by convex programming. \gls{SCP}-based methods have been widely applied across the aerospace domain, including in launch ascent guidance \citep{benedikterConvexApproachThreeDimensional2021, miaoSuccessiveConvexificationAscent2022}, entry-descent-landing \citep{acikmeseConvexProgrammingApproach2007, reynoldsDualQuaternionBasedPowered2020, kamathRealTimeSequentialConic2023, maReducedSpaceSequential2024}, and space missions employing conventional and low-thrust engines \citep{hofmannComputationalGuidanceLowThrust2023, kumagaiAdaptiveMeshSequentialConvex2024}. Applications of \gls{SCP} to solar sail trajectory design \citep{songSolarsailDeepSpace2019} have historically been limited by the difficulty of convexifying the solar sail control; however, recent work on the lossless convexification of solar sail dynamics \citep{oguriLosslessControlConvexFormulation2024} has enabled tractable implementations of solar sail trajectory design problems within \gls{SCP} frameworks.

To address the coupled challenges of solar sail trajectory design in \glspl{eLLO}, this paper presents a two-stage framework that enables efficient station-keeping by exploiting the translation dynamics of the eccentricity vector. The first stage, discussed in Section~\ref{sec_misocp_algorithm}, introduces a \gls{MISOCP} algorithm leveraging the translation theorem that identifies favorable orbital configurations for solar sail station-keeping as well as initial guesses for sail control profiles. This algorithm is loosely related to impulsive eccentricity vector control strategies \citep{wallaceLowLunarOrbit2012, sweetserDesignExtendedMission2012, yarndleyOriginsApplicationTranslation2026} but instead utilizes the solar sail to perform eccentricity vector translations throughout the trajectory. The second stage, discussed in Section~\ref{sec_scp_translation}, proposes a lightweight \gls{SCP} formulation that employs a lossless convexification of solar sail control \citep{oguriLosslessControlConvexFormulation2024} to efficiently refine the solar sail control profiles for the maneuvers identified in the first stage. Finally, Section~\ref{sec_results_discussion} demonstrates the effectiveness of the proposed approach through a scenario inspired by the \gls{LRO} mission \citep{beckmanStationkeepingLunarReconnaissance2007} using a solar sail design based on the NEA Scout mission \citep{lantoineTrajectoryManeuverDesign2024}.

\section{Background}
\label{sec_background}

In this section, we present the high-fidelity dynamical environment employed throughout this work. For \glspl{eLLO}, the central two-body gravity of the Moon is by far the strongest acceleration, followed by perturbations from the non-spherical lunar gravitational field. Third-body effects are nearly negligible at these altitudes \citep{foltaLunarFrozenOrbits2006}, but are included for completeness. A non-ideal flat-plate solar sail model is used to model the \gls{SRP} acceleration, and a conical shadow model is included to model the effects of eclipses by the Moon and Earth.

\subsection{Frame and state definitions}
\label{sec_frame_background}

The spacecraft dynamics are integrated within the inertial lunar mean equator-of-date J2000 reference frame, denoted by $F_\text{LME2000}$. In this frame, the $x$-axis points toward the intersection between the lunar equator of date and the J2000 equator, the $z$-axis points toward the lunar north pole of date J2000, and the $y$-axis completes the right-handed frame. The origin of the frame is located at the Moon's center of mass, which is retrieved from the DE440 ephemeris \citep{parkJPLPlanetaryLunar2021}. To use lunar spherical harmonic models, which rotate with the lunar surface, we introduce the rotating lunar principal-axis frame, denoted by $F_{\text{PA440}}$. This frame is defined within the DE440 ephemeris model \citep{parkJPLPlanetaryLunar2021}, and approximately rotates around the $z$-axis of $F_\text{LME2000}$ in a counterclockwise direction once every 27.32 days (the lunar sidereal period).

Generally, the spacecraft state vector is defined in Cartesian coordinates as $\boldsymbol{x} = [\boldsymbol{r}, \boldsymbol{v}, m]^T$, where $\boldsymbol{r}$ and $\boldsymbol{v}$ are the position and velocity vectors in the $F_\text{LME2000}$ frame, and $m$ is the spacecraft mass. For the \gls{SCP} formulation in Section~\ref{sec_scp_translation}, the \glspl{MEE} \citep{walkerSetModifiedEquinoctial1985} are used for the dynamics propagation due to their superior linearization properties and computational efficiency compared to other coordinate choices \citep{junkinsExplorationAlternativeState2019}. The \glspl{MEE} are defined in terms of the Keplerian orbital elements as follows,
\begin{equation}
\begin{aligned}
p &= a(1 - e^2) \\
f &= e \cos(\omega + \Omega) \\
g &= e \sin(\omega + \Omega) \\
h &= \tan\left(\frac{I}{2}\right) \cos(\Omega) \\
k &= \tan\left(\frac{I}{2}\right) \sin(\Omega) \\
L &= \Omega + \omega + \nu
\end{aligned}
\end{equation}
where $a$ is the semi-major axis, $e$ is the eccentricity, $I$ is the inclination, $\Omega$ is the longitude of the ascending node, $\omega$ is the argument of periapsis, and $\nu$ is the true anomaly.

\subsection{Dynamics}
\label{sec_dynamics}

To define the gravitational dynamics of the spacecraft in an extensible framework, the overall acceleration $\boldsymbol{a}$ is expressed as the sum of several components, including the gravitational accelerations from the Moon, Earth, and Sun, the non-spherical lunar gravity field, and the \gls{SRP} acceleration from the solar sail. Accordingly, the dynamics are expressed in the Cartesian $F_\text{LME2000}$ frame as follows,
\begin{equation} \label{equation_cartesian_dynamics}
\dot{\boldsymbol{x}} = \begin{bmatrix}\dot{\boldsymbol{r}}(t) \\ \dot{\boldsymbol{v}}(t) \\ \dot{m}(t) \end{bmatrix} = \begin{bmatrix}\boldsymbol{v} \\ \boldsymbol{a}_{\text{CB}} + \boldsymbol{a}_{\text{3B}} + \boldsymbol{a}_{\text{SHM}} + \boldsymbol{a}_{\text{SRP}} \\ 0 \end{bmatrix},
\end{equation}
where $\boldsymbol{a}_{\text{CB}}$ is the central two-body acceleration due to the Moon, $\boldsymbol{a}_{\text{3B}}$ is the acceleration due to third-body perturbations from the Earth and Sun, $\boldsymbol{a}_{\text{SHM}}$ is the acceleration due to the non-spherical lunar gravity field, and $\boldsymbol{a}_{\text{SRP}}$ is the acceleration due to \gls{SRP} on the solar sail.

\subsubsection{Gravitational dynamics}
\label{sec_gravitational_dynamics}

The central two-body acceleration is expressed as the standard Newtonian point-mass gravity,
\begin{equation}
\boldsymbol{a}_{\text{CB}} = -\frac{\mu_\text{M}}{r^3} \boldsymbol{r},
\end{equation}
where $\mu_\text{M}$ is the gravitational parameter of the Moon, and $r = \|\boldsymbol{r}\|$ is the distance of the spacecraft from the Moon. Similarly, the third-body perturbations due to the Earth and Sun are expressed as
\begin{equation}
\boldsymbol{a}_{\text{3B}} = \sum_{i\in \mathcal{B}} \mu_i \left( \frac{\boldsymbol{r}_i - \boldsymbol{r}}{\|\boldsymbol{r}_i - \boldsymbol{r}\|^3} - \frac{\boldsymbol{r}_i}{\|\boldsymbol{r}_i\|^3} \right),
\end{equation}
where $\mu_i$ is the gravitational parameter of the third body $i$, and $\boldsymbol{r}_i$ is the position vector of the third body $i$ with respect to the Moon. $\mathcal{B} = \{\text{Earth}, \text{Sun}\}$ is the set of third bodies considered. The positions of the Earth and Sun are time-dependent and are retrieved from the DE440 ephemeris \citep{parkJPLPlanetaryLunar2021}.

Perturbations due to the non-spherical lunar gravity are modeled using a spherical harmonic expansion of the gravitational potential (excluding the central two-body term, which is already included in $\boldsymbol{a}_{\text{CB}}$), and truncated to a maximum order and degree of $N=M=51$. The optimal choice of truncation for the model is highly dependent on the intended altitude of the orbit, but it is critical to consider since the computational cost of evaluating the spherical harmonic model scales with $O(N^2)$. This truncation is selected based on previous studies into the accuracy of the eccentricity-vector propagation \citep{yarndleyOriginsApplicationTranslation2026} at different truncation levels. For detailed mission analysis, it is likely that a higher order and degree may be required; however, in general, the resultant impact on the feasibility of the proposed algorithms would be minimal due to the limited changes in eccentricity-vector propagation. Therefore, the acceleration due to the non-spherical lunar gravity field is expressed as,
\begin{equation}
\boldsymbol{a}_{\text{SHM}} = \boldsymbol{C}_{\text{PA440} / \text{LME2000}} \frac{\partial U}{\partial \boldsymbol{r_\text{PA440}}},
\end{equation}
where $\boldsymbol{C}_{\text{PA440} / \text{LME2000}}$ is the time-dependent direction cosine matrix that transforms vectors from the $\text{F}_{\text{PA440}}$ frame to the $\text{F}_\text{LME2000}$ frame, retrieved from the DE440 ephemeris \citep{parkJPLPlanetaryLunar2021}, $\boldsymbol{r_\text{PA440}}=\boldsymbol{C}_{\text{LME2000} / \text{PA440}} \boldsymbol{r_\text{LME2000}}$ is the position vector of the spacecraft expressed in the $\text{F}_{\text{PA440}}$ frame, and $U$ is the spherical harmonic expansion of the gravitational potential expressed in the $\text{F}_{\text{PA440}}$ frame as,
\begin{equation}
U = \frac{\mu_\text{M}}{r} \sum_{n=2}^{N} \sum_{m=0}^{\min(n,M)} \left(\frac{R_\text{M}}{r}\right)^n \left( C_{nm} \cos m\lambda + S_{nm} \sin m\lambda \right) P_{nm}(\sin \phi),
\end{equation}
where $R_\text{M}$ is the mean radius of the Moon, $P_{nm}$ are the associated Legendre polynomials, $C_{nm}$ and $S_{nm}$ are the spherical harmonic coefficients, and $\phi$ and $\lambda$ are the latitude and longitude of the spacecraft in the $\text{F}_{\text{PA440}}$ frame. The spherical harmonic coefficients are retrieved from the GL0660B lunar gravity model \citep{konoplivJPLLunarGravity2013}, which is based on the data collected by NASA's \gls{GRAIL} mission.

\subsubsection{Solar sail dynamics}
\label{sec_solar_sail_dynamics}

A non-ideal flat-plate solar sail model is used to calculate the \gls{SRP} acceleration of the solar sail. A detailed description of this model is presented by \citet{oguriSolarSailingPrimer2022}. This model expresses the total \gls{SRP} acceleration acting on the sail as the sum of specular reflection, diffuse reflection, and absorption components, which either act along the normal or sunlight directions. The total \gls{SRP} acceleration, parameterized by the sail normal direction $\hat{\boldsymbol{u}}_n$, is therefore,
\begin{equation}\label{srp_flat_plate_acceleration}
\boldsymbol{a}_{\text{SRP}} = -\frac{CA}{m}\left(\frac{r_\oplus}{r}\right)^2 \left( \hat{\boldsymbol{u}}_n \cdot \hat{\boldsymbol{u}}_r \right) 
\left[ \underbrace{2\nu \,\hat{\boldsymbol{u}}_n}_{\text{diffuse}} + \underbrace{4\mu \left(\hat{\boldsymbol{u}}_n \cdot \hat{\boldsymbol{u}}_r  \right) \hat{\boldsymbol{u}}_n}_{\text{specular}} + \underbrace{\left(1 - 2\mu\right) \hat{\boldsymbol{u}}_r}_{\text{absorption}} \right],
\end{equation}
where $r_\oplus$ is the reference distance from the Sun (typically 1~AU), $C$ is the solar flux at $r_\oplus$, $A$ is the sail area, $m$ is the spacecraft mass, $r$ is the distance of the spacecraft from the Sun, $\hat{\boldsymbol{u}}_n$ is the unit normal vector of the sail, $\hat{\boldsymbol{u}}_r$ is the unit vector from the spacecraft to the Sun, $\mu$ is the solar sail specular reflection coefficient, and $\nu$ is the solar sail diffuse reflection coefficient. Combined, the parameters $\mu$ and $\nu$ define the relative proportions of absorption, specular reflection, and diffuse reflection that define the non-ideal flat-plate solar sail model.

\begin{figure}[h]
\centering
\includegraphics[width=3.25in]{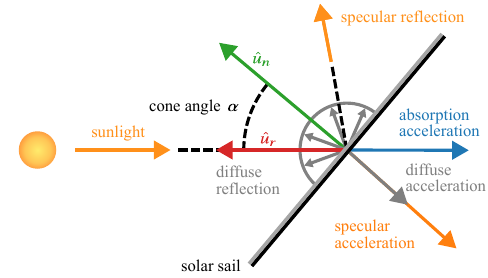}
\caption{\gls{SRP} acceleration components for the non-ideal flat-plate solar sail model relative to the sail normal and Sun line.}
\label{figure_srp_flat_plate}
\end{figure}

Fig.~\ref{figure_srp_flat_plate} illustrates the geometric construction of the non-ideal flat-plate solar sail model. Incoming sunlight is either diffusely reflected by the sail causing an acceleration in the opposite direction to the sail normal $\hat{\boldsymbol{u}}_n$, specularly reflected, again causing an acceleration in the opposite direction to the sail normal $\hat{\boldsymbol{u}}_n$, or absorbed by the sail, causing acceleration in the opposite direction to the sunlight $\hat{\boldsymbol{u}}_r$.

\begin{figure}[h]
\centering
\includegraphics[width=3.25in]{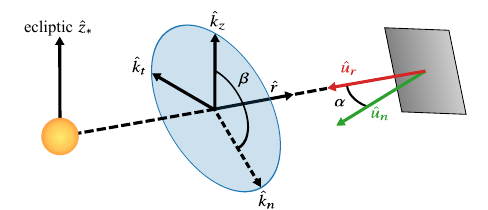}
\caption{Definition of the cone angle $\alpha$ and clock angle $\beta$ in the local Sun-pointing frame $F_\text{SN}$ used to parameterize the solar sail attitude.}
\label{figure_srp_frame}
\end{figure}

It is also customary to define the orientation of the solar sail in terms of the cone angle $\alpha$ and clock angle $\beta$. These angles are defined in a rotating frame $F_\text{SN}$ with respect to the Sun, as illustrated in Fig.~\ref{figure_srp_frame}. The cone angle $\alpha$ is defined as the angle between the sail normal $\hat{\boldsymbol{u}}_n$ and the sunlight direction $\hat{\boldsymbol{u}}_r$, and therefore determines the total magnitude of the \gls{SRP} acceleration. Conversely, the clock angle $\beta$ is defined as the angle between the projection of the sail normal onto the plane perpendicular to the sunlight direction and a reference direction in this plane and therefore determines the direction of the \gls{SRP} acceleration within this plane.

The transformation between the sail normal and the cone and clock angles is expressed as
\begin{subequations}
\begin{align}
    \cos \alpha &= \hat{\boldsymbol{u}}_n \cdot \hat{\boldsymbol{u}}_r \\
    \beta &= s \cos^{-1} \left( \hat{\boldsymbol{k}}_n \cdot \hat{\boldsymbol{k}}_z \right) \\
    \hat{\boldsymbol{k}}_n &= \left[ \hat{\boldsymbol{u}}_n - \left( \hat{\boldsymbol{u}}_n \cdot \hat{\boldsymbol{u}}_r \right) \hat{\boldsymbol{u}}_r \right] / \,\| \hat{\boldsymbol{u}}_n - \left( \hat{\boldsymbol{u}}_n \cdot \hat{\boldsymbol{u}}_r \right) \hat{\boldsymbol{u}}_r \| \\
    \hat{\boldsymbol{k}}_z &= \left[ \hat{\boldsymbol{z}}^* - \left( \hat{\boldsymbol{z}}^* \cdot \hat{\boldsymbol{u}}_r \right) \hat{\boldsymbol{u}}_r \right] / \,\| \hat{\boldsymbol{z}}^* - \left( \hat{\boldsymbol{z}}^* \cdot \hat{\boldsymbol{u}}_r \right) \hat{\boldsymbol{u}}_r \| \label{equation_srp_kz}\\
    s &= \text{sign} \left[ - \left( \hat{\boldsymbol{k}}_n \times \hat{\boldsymbol{k}}_z \right) \cdot \hat{\boldsymbol{u}}_r \right]
\end{align}
\end{subequations}

The sail normal is expressed in the rotating frame $F_\text{SN} = ( \hat{\boldsymbol{r}}, \hat{\boldsymbol{k}}_t, \hat{\boldsymbol{k}}_z )$ where $\hat{\boldsymbol{r}}$ is the unit vector from the Sun to the spacecraft, $\hat{\boldsymbol{k}}_z$ is defined as in Equation~\ref{equation_srp_kz} and $\hat{\boldsymbol{k}}_t = \hat{\boldsymbol{r}} \times \hat{\boldsymbol{k}}_z$ completes the right-handed frame. By definition, $\hat{\boldsymbol{r}} = -\hat{\boldsymbol{u}}_r$, so $\hat{\boldsymbol{u}}_r$ denotes the spacecraft-to-Sun direction used in the flat-plate model, whereas $\hat{\boldsymbol{r}}$ denotes the opposite Sun-to-spacecraft direction used to define $F_\text{SN}$. Then, the sail unit normal $\hat{\boldsymbol{u}}_n$ is expressed in the rotating frame $F_\text{SN}$ as a function of the cone angle $\alpha$ and clock angle $\beta$,
\begin{equation}
\hat{\boldsymbol{u}}_n = \begin{bmatrix} -\cos \alpha \\  \sin \alpha \sin \beta \\  \sin \alpha \cos \beta \end{bmatrix}.
\end{equation}

These angles and the geometric construction of the unit vectors are illustrated in Fig.~\ref{figure_srp_frame}. The cone angle is usually restricted to the range $\alpha \in [0, \pi/2]$ since negative cone angles would cause the sail to face away from the Sun. Furthermore, in the case of some missions, such as NEA Scout \citep{lantoineTrajectoryManeuverDesign2024}, the cone angle is further restricted due to power considerations. There are typically no constraints on the clock angle.

\begin{figure}[h]
\centering
\includegraphics[width=3.25in]{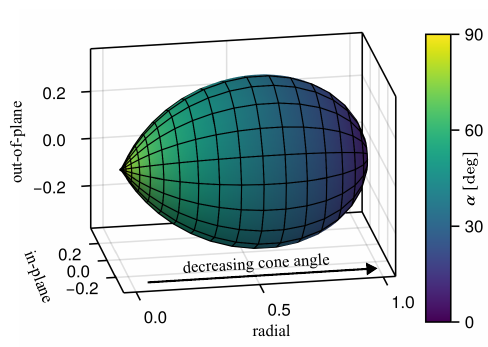}
\caption{Normalized feasible \gls{SRP} acceleration set for a representative non-ideal flat-plate solar sail with NEA Scout-like optical properties.}
\label{figure_srp_acceleration}
\end{figure}

A visualization of the feasible set of \gls{SRP} accelerations for a non-ideal flat-plate solar sail is shown in Fig.~\ref{figure_srp_acceleration}. This figure shows the normalized set of feasible accelerations for a solar sail with $\mu=0.40495$ and $\nu=0.014957$, which are representative values for current solar sail technology based on the NEA Scout mission \citep{oguriSolarSailingPrimer2022}. The region is symmetric about the Sun-line axis $\hat{\boldsymbol{r}}=-\hat{\boldsymbol{u}}_r$, forming a surface in $\mathbb{R}^3$. Since only a surface of accelerations is formed, it is much more difficult to work with solar sail control when compared with conventional propulsion systems, which can typically provide acceleration in any direction and over a range of magnitudes forming a volume in $\mathbb{R}^3$.

\subsection{Shadowing effects}
\label{sec_shadowing_effects}

A conical shadow model (also known as an umbra-penumbra model) is used to represent the effects of eclipses by the Moon and Earth on the \gls{SRP} acceleration of the solar sail \citep{hubauxSymplecticIntegrationSpace2012}. This model approximates the umbra and penumbra regions as two cones extending from the eclipsing body, and computes the fraction of sunlight incident on the sail based on its position within these cones. Compared to the simpler cylindrical shadow model \citep{bryantEffectSolarRadiation1961}, this model provides a continuous variation in the \gls{SRP} acceleration during eclipse transitions, which improves stability during numerical integration.

\begin{figure}[h]
\centering
\includegraphics[width=3.25in]{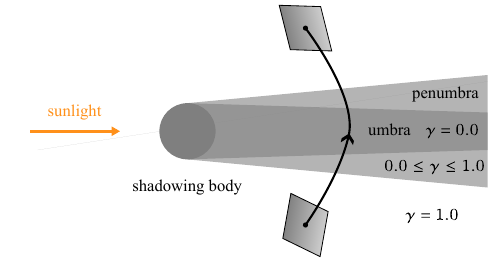}
\caption{Geometry of the conical umbra-penumbra shadow model used to scale the solar-sail \gls{SRP} acceleration by the shadow factor $\gamma$.}
\label{figure_srp_shadowing}
\end{figure}

The geometric construction of the conical shadow model is illustrated in Fig.~\ref{figure_srp_shadowing}. The shadow factor $\gamma$ defines the fraction of sunlight incident on the sail, where $\gamma = 1$ corresponds to full sunlight and $\gamma = 0$ corresponds to a full eclipse (umbra). The \gls{SRP} acceleration is then modified to include the shadow factor as
\begin{equation}
\boldsymbol{a}_{\text{SRP}} = -\gamma \frac{CA}{m}\left(\frac{r_\oplus}{r}\right)^2 
\left( \hat{\boldsymbol{u}}_n \cdot \hat{\boldsymbol{u}}_r \right)
\left[ 
2\nu\,\hat{\boldsymbol{u}}_n 
+ 4\mu \left(\hat{\boldsymbol{u}}_n \cdot \hat{\boldsymbol{u}}_r\right)\hat{\boldsymbol{u}}_n 
+ (1 - 2\mu)\hat{\boldsymbol{u}}_r 
\right].
\end{equation}

To compute the shadow factor $\gamma$, the solar angular size $\alpha_\odot$ and apparent area of the solar disk $A_\odot$ are first determined, as viewed from the spacecraft. These are defined as
\begin{equation}
\begin{aligned}
\alpha_\odot &= \arcsin \left( \frac{R_\odot}{\|\boldsymbol{r}_{s/sc}\|} \right), \\
A_\odot &= \pi\alpha_\odot^2,
\end{aligned}
\end{equation}
where $R_\odot$ is the radius of the Sun, and $\boldsymbol{r}_{s/sc}$ is the position vector of the Sun relative to the spacecraft. Next, for each shadowing body $b$ (in this case, the Moon and Earth), the angular separation $\theta_b$ between the Sun and the eclipsing body, as viewed from the spacecraft, is computed, along with the angular size of the eclipsing body $\alpha_b$. These are defined as,
\begin{equation}
\begin{aligned}
\theta_b &= \arccos \left( 
\frac{\mathbf{r}_{s/sc}\cdot\mathbf{r}_{b/sc}}{\|\mathbf{r}_{s/sc}\| \|\mathbf{r}_{b/sc}\|}
\right), \\
\alpha_b &= \arcsin \left(\frac{R_b}{\|\mathbf{r}_{b/sc}\|}\right),
\end{aligned}
\end{equation}
where $R_b$ is the radius of the eclipsing body $b$, and $\boldsymbol{r}_{b/sc}$ is the position vector of the eclipsing body $b$ relative to the spacecraft. Then, for each shadowing body $b$, the shadow factor $\gamma_b$ is computed using the angular separation $\theta_b$, solar angular size $\alpha_\odot$, and eclipsing body angular size $\alpha_b$,
\begin{equation}
\gamma_b = \begin{cases}
0, & \theta_b \le |\alpha_\odot - \alpha_b| \\
1, & \theta_b \ge \alpha_\odot + \alpha_b \\
1 - \dfrac{A(\alpha_\odot,\alpha_b,\theta_b)}{A_\odot}, & \text{otherwise}
\end{cases}
\end{equation}
where the partial overlap area $A(\alpha_\odot,\alpha_b,\theta_b)$ between two circular regions is
\begin{equation}
\begin{aligned}
A(\alpha_\odot,\alpha_b,\theta_b) =\;&
\alpha_\odot^2 \arccos \left(
\frac{\theta_b^2+\alpha_\odot^2-\alpha_b^2}{2\theta_b\alpha_\odot}
\right)
+
\alpha_b^2 \arccos \left(
\frac{\theta_b^2+\alpha_b^2-\alpha_\odot^2}{2\theta_b\alpha_b}
\right)
\\
&-
\frac{1}{2}\sqrt{
(-\theta_b + \alpha_\odot + \alpha_b)
(\theta_b + \alpha_\odot - \alpha_b)
(\theta_b - \alpha_\odot + \alpha_b)
(\theta_b + \alpha_\odot + \alpha_b)
}.
\end{aligned}
\end{equation}

Together, these formulas ensure that if the spacecraft is fully in the umbra of the eclipsing body $b$, then $\gamma_b = 0$; if it is fully illuminated, then $\gamma_b = 1$; and if it is in the penumbra region, then $0 < \gamma_b < 1$ based on the fraction of the solar disk that is obscured by the eclipsing body. Finally, an efficient approximation of the overall shadow factor $\gamma$ is obtained as the minimum of the individual shadow factors,
\begin{equation}
\gamma = \min_{b} \left( \gamma_b \right).
\end{equation}

\begin{figure}[h]
\centering
\includegraphics[width=3.25in]{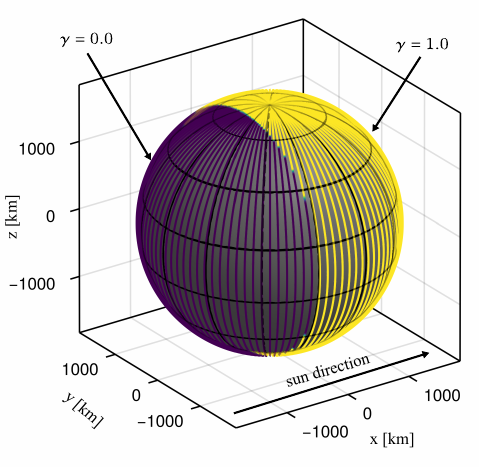}
\caption{Moon-shadow geometry for several 50-km polar lunar orbits, illustrating the short penumbra transition and frequent eclipse encounters expected in eLLO.}
\label{figure_llo_shadowing}
\end{figure}

Fig.~\ref{figure_llo_shadowing} illustrates the geometry of the shadow model for several 50~km polar lunar orbits. The umbra and penumbra regions of the Moon's shadow are visible, but the penumbra transition is abrupt due to the low altitude. It is also apparent that these orbits experience frequent shadowing, indicating the significant effect the shadowing model has on the solar sail \gls{SRP} acceleration, so it should be carefully accounted for during station-keeping.

\subsection{Full dynamical environment}
\label{sec_dynamical_environment}

In the dynamical integration, the equations of motion are non-dimensionalized in order to improve numerical stability. The distance unit (DU) is set to the mean radius of the Moon, $\text{DU}=R_\text{M} = 1737.4$~km, and the gravitational parameter of the Moon is set to $\mu_\text{M} = \text{DU}^3/\text{TU}^2 = 4902.80\ \text{km}^3/\text{s}^2$. Therefore, the time unit (TU) is derived as $\text{TU} = \sqrt{R_\text{M}^3/\mu_\text{M}} = 1034.26\ \text{s}$ and the velocity unit is thus VU = DU/TU = 0.8319~km/s. The gravitational parameters of the Earth and Sun are similarly rescaled to $\mu_\text{E} = 81.300569$~DU$^3$/TU$^2$ and $\mu_\odot = 888611.243$~DU$^3$/TU$^2$. All initial conditions are converted into these non-dimensional units, and the equations of motion are integrated in these units. Finally, the state is converted back to dimensional units for output.

\begin{figure}[h]
\centering
\includegraphics[width=\textwidth]{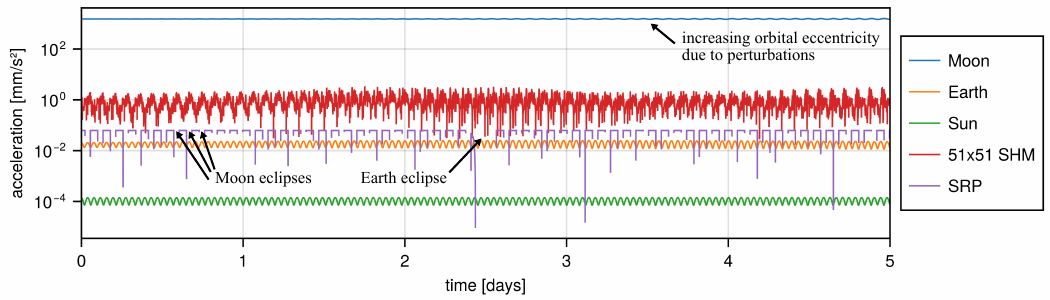}
\caption{Modeled acceleration magnitudes over 5 days for a 50-km polar lunar orbit, showing the relative scales of central gravity, non-spherical lunar gravity, solar-sail \gls{SRP}, and third-body perturbations.}
\label{figure_acceleration_llo}
\end{figure}

Fig.~\ref{figure_acceleration_llo} presents the relative magnitudes of the various modeled accelerations over a 5-day period for a 50~km altitude polar lunar orbit. The central two-body acceleration from the Moon is by far the strongest acceleration, followed by the non-spherical lunar gravity field perturbations, which are approximately three orders of magnitude smaller. The \gls{SRP} acceleration from the solar sail (assuming parameters similar to NEA Scout) is smaller still, but remains significant enough to be used for station-keeping maneuvers. Finally, the third-body perturbations from the Earth and Sun are even smaller and are therefore likely negligible at this altitude. These results support the findings of \citet{foltaLunarFrozenOrbits2006}, who state that third-body perturbations can be considered as negligible within \glspl{eLLO}.

\section{Control Strategy}
\label{sec_control_strategy}

\begin{figure}[h]
\centering
\includegraphics[width=\textwidth]{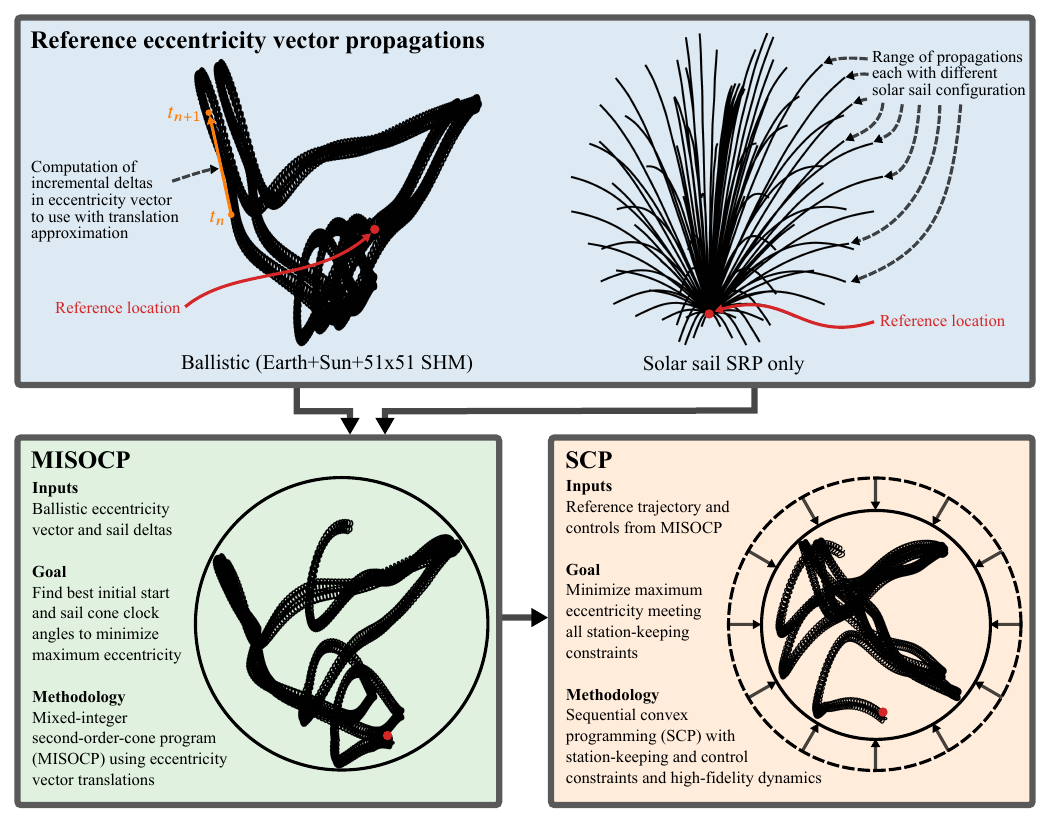}
\caption{Two-stage station-keeping workflow: translation-based \gls{MISOCP} search followed by high-fidelity \gls{SCP} refinement.}
\label{figure_translation_sail_explainer}
\end{figure}

This section presents the control strategy for maintaining the spacecraft within a defined station-keeping region using a solar sail. Fig.~\ref{figure_translation_sail_explainer} provides an overview of the main components and steps of the proposed control strategy. The first part of the strategy (the first two panels) seeks to find the optimal locations in the eccentricity-vector plane that minimize the maximum eccentricity, thereby identifying station-keeping strategies that tightly constrain the eccentricity vector. Additionally, the solar sail control is implemented through the inclusion of additional translations of the eccentricity vector, based on the permitted directions and magnitudes of the solar sail \gls{SRP} acceleration. This provides a suitable initial guess for the second part, namely the initialization of a constrained \gls{SCP} formulation that minimizes the cone angle of the solar sail and the maximum eccentricity, while ensuring the trajectory fully satisfies the dynamics and station-keeping constraints.

\subsection{Translation theorem with solar sailing}
\label{sec_translation_sailing}

The `translation theorem' describes an interesting behavior observed in the propagation of the eccentricity vector under perturbations from a non-spherical gravity field \citep{beckmanStationkeepingLunarReconnaissance2007}. In particular, the propagation of the eccentricity vector from any given starting location in the eccentricity-vector plane can be approximated as a translation of the eccentricity-vector propagation from a reference location. It applies as long as the translation is small, which makes it ideally suited to the problem of station-keeping, since the goal is to keep an orbit close to a nominal orbit. This can be exploited to efficiently search for optimal initial conditions in the eccentricity-vector plane that maximize the time before station-keeping constraints are violated \citep{yarndleyOriginsApplicationTranslation2026}. The efficiency of this approach stems from the fact that only a single numerical propagation is required from the reference location, which is particularly advantageous when working with high-order spherical harmonic models, where numerical integration is computationally expensive.

\begin{figure}[h]
\centering
\includegraphics[width=\textwidth]{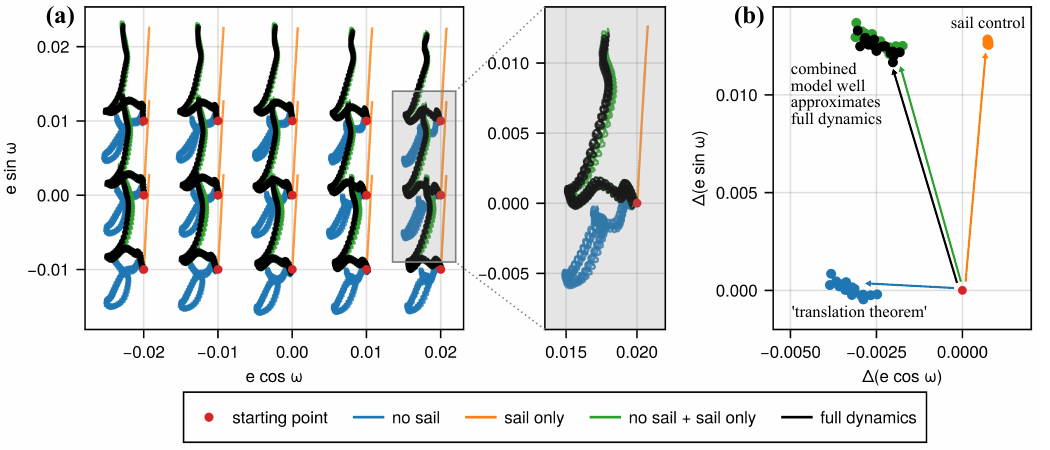}
\caption{Short eccentricity-vector propagations from multiple initial conditions, with panel (b) showing the relative changes used to assess the translation approximation.}
\label{figure_translation_control}
\end{figure}

The propagation of the eccentricity vector due to solar sailing exhibits a similar translation behavior. This observation enables the effects from both the non-spherical gravity field and solar sailing to be represented as separate translations of the eccentricity vector from a reference location, which is illustrated in Fig.~\ref{figure_translation_control}. In particular, (a) demonstrates the propagation of the eccentricity vector from a grid of starting locations, while (b) shows the relative changes in the eccentricity vector. The strong clustering of relative changes across starting locations in (b) indicates that the translations are largely independent of the starting point, and the agreement between the full and combined translations validates the separated approach.

\subsection{Translation-based MISOCP station-keeping algorithm}
\label{sec_misocp_algorithm}

Unlike previous studies, which relied on impulsive maneuvers to translate the eccentricity vector between optimal starting conditions, this work proposes a strategy that instead uses solar sailing maneuvers continuously throughout the trajectory. This makes the problem significantly more challenging due to the limited control actuation of solar sails. In contrast to previous work \citep{yarndleyOriginsApplicationTranslation2026}, the key difference in this strategy is the replacement of the greedy impulsive maneuver algorithm with a \gls{MISOCP} formulation that incorporates a limited set of feasible solar sail configurations at each time step. This formulation minimizes the maximum eccentricity encountered over the station-keeping duration. Compared to maximizing the time before violating a fixed eccentricity constraint, minimizing maximum eccentricity is more flexible. This prevents infeasibility problems and provides a clear estimate of the feasibility margin for a given trajectory.

The first step of the approach is to define a small set $\mathcal{C}$ of possible solar sail cone and clock angles $(\alpha, \beta)$ that can be selected for station-keeping maneuvers. The set is kept relatively small to limit the number of decision variables, since each configuration introduces binary variables in the \gls{MISOCP}. In this work, the cone angle is discretized into 10 equally spaced values from $0^\circ$ to $75^\circ$ and the clock angle into 10 equally spaced values from $0^\circ$ to $360^\circ$, resulting in 100 candidate configurations. A segmentation is then defined across the station-keeping duration, with index $n=1, 2, \dots, N$ and corresponding times $t_1, t_2, \dots, t_N$. The segmentation defines the times at which the eccentricity vector is checked, when translations to the eccentricity vector are applied, and when the solar sail configuration may be changed.

Next, the reference ballistic propagation without the solar sail is computed from the nominal (circular) station-keeping orbit. During this propagation, at each time $t_n$, the eccentricity is reset to zero in order to prevent any long-term drift, which would otherwise degrade the accuracy of the translation theorem approximation. The eccentricity vector, defined as $(C, S) = (e \cos \omega, e \sin \omega)$, is computed and stored at each $t_n$ and then the step-to-step changes $(\Delta C_n, \Delta S_n) = (C_{n+1} - C_{n}, S_{n+1} - S_n)$ are computed. These changes define the baseline evolution of the eccentricity vector near the origin in the absence of solar sailing.

In a similar manner, for each solar sail configuration and time $t_n$, a propagation including only the central two-body acceleration and the solar sail \gls{SRP} acceleration is performed using the initial conditions taken from the ballistic propagation. The corresponding eccentricity vectors are then computed, from which the step-to-step differences $(\Delta C_{n, (\alpha, \beta)}, \Delta S_{n, (\alpha, \beta)})$ are obtained. These represent the eccentricity vector translations that each solar sail configuration can achieve across each segment.

The \gls{MISOCP} formulation is then constructed. The decision variables consist of the eccentricity vector at each time step $(C_n, S_n)$, binary selection variables $s_{n, (\alpha, \beta)}$ indicating which solar sail configuration is chosen at each $t_n$, and an auxiliary variable $e_{\text{max}}$ that defines the maximum eccentricity over the entire trajectory. First, a linear expression for the propagation of the eccentricity vector across each segment is defined using precomputed translations,
\begin{align} \label{equation_eccentricity_dynamics_linear}
    (C_{n+1}, S_{n+1}) &= (C_n, S_n) + (\Delta C_n, \Delta S_n) + \sum_{(\alpha, \beta) \in \mathcal{C}} s_{n, (\alpha, \beta)} (\Delta C_{n, (\alpha,  \beta)}, \Delta S_{n, (\alpha, \beta)}).
\end{align}
Then, the binary variables are constrained such that exactly one solar sail configuration is selected at each segment,
\begin{align}
    \sum_{(\alpha, \beta) \in \mathcal{C}} s_{n, (\alpha, \beta)} &= 1 \label{equation_sail_angle_binary}, \\
    s_{n, (\alpha, \beta)} &\in \{0, 1\} \label{equation_sail_angle_binary_set}.
\end{align}
Finally, the maximum eccentricity variable $e_{\text{max}}$ is constrained to be greater than or equal to the magnitude of the eccentricity vector (the eccentricity) at each segment. This is imposed using \gls{SOC} to obtain the $L_2$ norm,
\begin{align} \label{equation_eccentricity_maximum}
    e_{\text{max}} \geq \left\|(C_n, S_n) \right\|_2\quad \text{(SOC)}.
\end{align}
The objective is to minimize the maximum eccentricity $e_{\text{max}}$. The complete \gls{MISOCP} formulation is therefore
\begin{mini*}
    {}{e_{\text{max}}\quad\text{(maximum eccentricity)}}{}{}
    \addConstraint{\eqref{equation_eccentricity_dynamics_linear}}{}{\quad\text{(eccentricity vector propagation)}}
    \addConstraint{\eqref{equation_sail_angle_binary}}{}{\quad\text{(single sail configuration selected)}}
    \addConstraint{\eqref{equation_sail_angle_binary_set}}{}{\quad\text{(binary sail configuration variables)}}
    \addConstraint{\eqref{equation_eccentricity_maximum}}{}{\quad\text{(maximum eccentricity calculation)}}.
\end{mini*}

This \gls{MISOCP} contains only \gls{SOC} and linear constraints and has a linear objective, so is efficiently solvable using modern conic solvers that support mixed-integer programming. Although the introduction of binary variables introduces computational complexity to the formulation, the number of configurations is kept relatively small, and the structure of the problem is highly exploitable, so the problem remains tractable. In terms of implementation, the \texttt{JuMP.jl} \citep{lubinJuMPRecentImprovements2023} modeling language is used to construct the \gls{MISOCP} formulation, and Gurobi \citep{gurobi2025} is used as the solver.

When applying the eccentricity vector station-keeping strategy described above, there are four remaining parameters for the mission designer to consider when planning the station-keeping. These are the initial epoch, the semi-major axis $a$, the inclination $i$, and the longitude of the ascending node $\Omega$. The choice of the fast variable (e.g., the initial true anomaly) does not significantly affect the solution. Appropriate values for these parameters must be selected in accordance with the mission goals and operational requirements.

\subsection{SCP formulation for eccentricity vector translation}
\label{sec_scp_translation}

To assess the feasibility of the translation sequences generated by the \gls{MISOCP} algorithm, a convex optimization approach based on \gls{SCP} is implemented. This computes the optimal cone and clock angles for the solar sail to best satisfy the station-keeping constraints. When compared to traditional indirect or direct methods for trajectory optimization, \gls{SCP} has the advantage of being able to efficiently solve problems with complex dynamics and constraints by iteratively solving a series of convex subproblems. This makes it particularly well-suited for difficult problems, such as those involving solar sails, and, importantly, it is computationally efficient, especially when handling the high-fidelity non-spherical gravity models required near the lunar surface.

Following the general approach of \gls{SCP} \citep{malyutaConvexOptimizationTrajectory2022}, the first step is to find an appropriate linearization of the dynamical system with solar sailing included. This linearization is derived from the spacecraft dynamics in Eq.~\eqref{equation_cartesian_dynamics}, which depend on the state vector, time, and the cone and clock angles of the solar sail. The trajectory is then divided into multiple segments, indexed $n= 1, 2, ..., N$, as in a direct transcription method. The number of segments $N$ is chosen based on an intended time interval between sail control updates. In most of this work, 1-day intervals are used, but these can be easily updated depending on the scenario.

However, the direct use of cone and clock angles as control variables typically results in poor linearization performance. Therefore, in this work, the control variables are instead taken to be the \gls{SRP} acceleration vectors in the rotating frame $F_{SN}$, normalized such that the magnitude is unity when the solar sail directly faces the Sun. Then, convexified constraints are introduced to ensure that the resulting acceleration remains within the feasible set of solar sail accelerations, as described in Section~\ref{sec_solar_sail_dynamics}. This approach was first introduced by \citet{oguriLosslessControlConvexFormulation2024} and is adapted here for the station-keeping problem.

Using these dynamics and discretization, the linearized dynamic constraints are constructed around a reference trajectory. The \gls{SCP} process requires an appropriate initial guess for the reference trajectory, which should be as close as possible in state to the optimal trajectory. Generally, a purely ballistic reference trajectory is sufficient for station-keeping; however, in this work, a very good reference trajectory is provided by the \gls{MISOCP} algorithm. Each segment is assigned a single control vector $\boldsymbol{u}_n=(u_{n,x}, u_{n,y}, u_{n,z})$. Then, given the reference trajectory ($\boldsymbol{\bar{x}_n}, \boldsymbol{\bar{u}}_n$), the dynamics, and the segmentation $n = 1, 2, ..., N$, a discrete form of the spacecraft dynamics is obtained and used as a convex constraint,
\begin{align} \label{equation_dynamics_linear}
    \boldsymbol{x}_{n+1} = \boldsymbol{A}_n (\boldsymbol{x}_n - \boldsymbol{\bar{x}}_n) + \boldsymbol{B}_n (\boldsymbol{u}_n - \boldsymbol{\bar{u}}_n) + \boldsymbol{\bar{x}}_{n+1} + \boldsymbol{\sigma}_n,
\end{align}
where $\boldsymbol{\sigma}_n$ are virtual controls used to ensure the feasibility of the linearized dynamics and,
\begin{align}
    \boldsymbol{A}_n &= \left. \left[\frac{\partial}{\partial \boldsymbol{x}} \int_{t_n}^{t_{n+1}}\dot{\boldsymbol{x}}\,\text{d}t \right]\right|_{(\boldsymbol{\bar{x}}_n, \boldsymbol{\bar{u}}_n)}\\
    \boldsymbol{B}_n &= \left. \left[\frac{\partial}{\partial \boldsymbol{u}} \int_{t_n}^{t_{n+1}} \dot{\boldsymbol{x}}\,\text{d}t \right]\right|_{(\boldsymbol{\bar{x}}_n, \boldsymbol{\bar{u}}_n)}
\end{align}

The matrix $\boldsymbol{A}_n$ is the state transition matrix (STM), representing the sensitivity of the final state $\boldsymbol{x}_{n+1}$ of each segment with respect to the initial state $\boldsymbol{x}_{n}$. Similarly, $\boldsymbol{B}_n$ represents the sensitivity of the final state with respect to the solar sail acceleration $\boldsymbol{u}_n$. Importantly, in this formulation, the sail dynamics are expressed in terms of the acceleration $\boldsymbol{u}_n$ rather than the cone and clock angles.

Instead of relying on analytic expressions, the partial derivatives for $\boldsymbol{A}_n$ and $\boldsymbol{B}_n$ are computed with \gls{AD}. This is applied directly to the initial conditions of a numerical integration solver, in this case, \texttt{Vern9} from the \texttt{DifferentialEquations.jl} \citep{rackauckasDifferentialEquationsJlPerformant2017} library with absolute and relative tolerances of $10^{-10}$. The \gls{AD} is computed in forward mode via \texttt{ForwardDiff.jl} \citep{revelsForwardModeAutomaticDifferentiation2016}, interfaced through \texttt{DifferentiationInterface.jl} \citep{dalle2025commoninterfaceautomaticdifferentiation}.

The use of Cartesian elements in the state vector $\boldsymbol{x}_n$ can lead to convergence difficulties in \gls{SCP}, particularly for trajectories with many revolutions, such as is the case in this work. To address this, the state is instead expressed in \gls{MEE} rather than Cartesian elements. Rather than rewriting the dynamical equations, which is generally very complex, the coordinate transformation is embedded directly within the dynamics. The dynamical function $\dot{\boldsymbol{x}} = f(\boldsymbol{x}, t, \boldsymbol{u})$ is modified such that the input state is first converted from \gls{MEE} to Cartesian, the Cartesian dynamics evaluated, and the resulting Cartesian derivatives are then mapped back to \gls{MEE} using the Jacobian of the transformation. Although this process introduces some computational overhead, it significantly improves the convergence behavior of the \gls{SCP} process while still allowing derivatives to be expressed in Cartesian form.
\begin{align}
    \dot{\boldsymbol{x}}_{\text{MEE}} &= \left.\frac{\partial \boldsymbol{x}_{\text{MEE}}}{\partial \boldsymbol{x}_{\text{cart}}} \right|_{\boldsymbol{x}_{\text{cart}}} \dot{\boldsymbol{x}}_{\text{cart}}
\end{align}

\begin{figure}[h]
\centering
\includegraphics[width=3.25in]{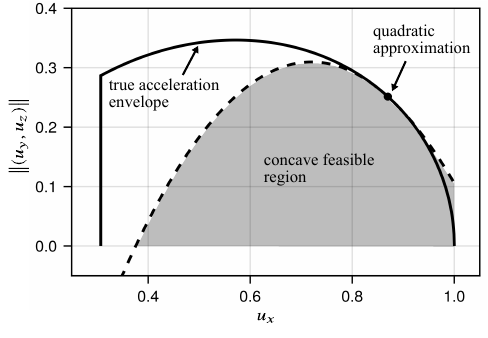}
\caption{Geometric construction of the lossless control-convex constraint used in the \gls{SCP} formulation to bound feasible solar-sail accelerations.}
\label{figure_lossless_sail_explainer}
\end{figure}

Next, the control vector $\boldsymbol{u}_n$ must be constrained to ensure that the resulting solar sail acceleration remains within the feasible set of accelerations. Following the approach of \citet{oguriLosslessControlConvexFormulation2024}, this feasible set is represented using a sequence of constraints that form an approximation of the convex hull of possible solar sail accelerations. This is visualized in Fig.~\ref{figure_lossless_sail_explainer}. At the optimal solution, these constraints are driven to binding by the introduction of a penalty term on $u_x$ in the objective function, which is added to implicitly enforce minimization of the cone angle. The resulting set of constraints is expressed as
\begin{align}
    \gamma(\alpha_{\text{max}}) &\leq u_{n,x} \leq \gamma(\alpha_{\text{min}}), \label{equation_sail_angle_limits} \\
    u_{n,yz} &\geq \left\| (u_{n,y}, u_{n,z})\right\|_2 \quad (\text{SOC}), \label{equation_sail_transverse_norm}\\
    u_{n,yz} &\leq \frac{1}{2} \left. \frac{\partial^2 h}{\partial u_x^2} \right|_{\bar{u}_{n,x}} (u_{n, x} - \bar{u}_{n, x})^2 + \left. \frac{\partial h}{\partial u_x} \right|_{\bar{u}_{n,x}} (u_{n, x} - \bar{u}_{n, x}) + h(\bar{u}_{n,x}) \quad (\text{SOC}) \label{equation_sail_transverse_taylor},
\end{align}
where,
\begin{align}
    C_1 &= 4 \mu, \\
    C_2 &= 2 \nu, \\
    C_3 &= 1 - 2 \mu, \\
    a &= \frac{(3 C_1 C_3 - C_2^2)}{3 C_1^2}, \\
    b &= \frac{(2 C_2^3 - 9 C_1 C_2 C_3 - 27 C_1^2 u_x)}{27 C_1^3}, \\
    \gamma(\alpha) &= C_1 \cos^3(\alpha) + C_2 \cos^2(\alpha) + C_3 \cos(\alpha), \\
    \tau(u_x) &= -2 \sqrt{\frac{a}{3}} \sinh\left[\frac{1}{3} \sinh^{-1}\left(\frac{3 b}{2 a}\sqrt{\frac{3}{a}}\right)\right] - \frac{C_2}{3 C_1}, \\
    h(u_x) &= (C_1 \tau + C_2) \tau \sqrt{1 - \tau^2}.
\end{align}

Firstly, the radial acceleration $u_{n,x}$ is bounded between the minimum and maximum feasible values, which are determined by the cone angle limits $\alpha_{\text{min}}$ and $\alpha_{\text{max}}$. In this work, $\alpha_{\text{min}} = 0.95^\circ$ and $\alpha_{\text{max}} = 75^\circ$. The non-zero value for $\alpha_{\text{min}}$ prevents numerical instabilities arising from the derivative approximation (see the region near $u_x = 1.0$ in Fig.~\ref{figure_lossless_sail_explainer}) and does not significantly affect the feasibility or optimality of the problem. Secondly, a \gls{SOC} constraint is applied to compute the norm of the transverse acceleration components $u_{n,y}$ and $u_{n,z}$. Finally, a Taylor series expansion of the feasible set boundary is used to provide an upper bound on the transverse acceleration components, which is updated at each \gls{SCP} iteration based on the reference control $\bar{u}_{n,x}$. The derivatives required for this Taylor series expansion are computed using \gls{AD}. For most realistic solar sail parameters, including those considered in this work, $h$ is a concave function \citep{oguriLosslessControlConvexFormulation2024}, so the Taylor series expansion forms a valid upper bound that can be expressed as a \gls{SOC} constraint despite the presence of quadratic terms.

Hard trust region constraints are introduced on the state to ensure that the dynamical linearization remains accurate. The trust regions are selected to have a constant size that remains unchanged as the \gls{SCP} algorithm progresses. This constraint is expressed as
\begin{align} \label{equation_state_trust_regions}
    -\epsilon \leq \boldsymbol{x}_n -\bar{\boldsymbol{x}}_n \leq \epsilon.
\end{align}
A value of $\epsilon = 1$ was found to work well, especially when leveraging \gls{MEE} elements. This primarily constrains changes to the angular component $L$ of the \gls{MEE} elements. Once the \gls{SCP} process has converged, smaller trust region sizes may be applied for several iterations to obtain higher-accuracy solutions if required.

Next, a linearization to obtain the eccentricity at each node of the trajectory is performed in order to obtain the maximum eccentricity $e_{\text{max}}$. The function $q(\boldsymbol{x})$ is defined as the conversion of state to eccentricity magnitude. This constraint is then expressed as
\begin{align} \label{equation_eccentricity_constraint}
    \left. \frac{\partial q}{\partial \boldsymbol{x}} \right|_{\boldsymbol{\bar{x}}_n} (\boldsymbol{x}_n - \boldsymbol{\bar{x}}_n) + q(\boldsymbol{\bar{x}}_n) \leq e_{\text{max}}.
\end{align}
By minimizing $e_{\text{max}}$, the trajectory is gradually guided to best satisfy the station-keeping region constraints while maintaining the feasibility of the subproblems throughout the \gls{SCP} iterations.

It is also important to control the semi-major axis drift across the trajectory. The semi-major axis is commonly restricted to a permissible band of values. Therefore, a similar linearization approach is used, where the function $r(\boldsymbol{x})$ computes the semi-major axis from the state. The constraint is then expressed as
\begin{align} \label{equation_semi_major_axis_constraint}
    \left. a_{\text{min}} + \xi_n \leq \frac{\partial r}{\partial \boldsymbol{x}} \right|_{\boldsymbol{\bar{x}}_n} (\boldsymbol{x}_n - \boldsymbol{\bar{x}}_n) + r(\boldsymbol{\bar{x}}_n) \leq a_{\text{max}} + \xi_n.
\end{align}
A virtual control $\xi_n$ is introduced to ensure feasibility and is penalized in the objective function. No explicit initial or final state constraints are applied, since the objective is simply to remain within the station-keeping region throughout the entire trajectory.

The main objective is then to minimize the magnitude of the virtual controls (ensuring feasibility when all are zero), minimize the cone angle of the solar sail, and minimize the maximum eccentricity $e_{\text{max}}$. The complete objective to be minimized is thus
\begin{align} \label{equation_scp_objective}
    J = e_{\text{max}} - 10^{-2} \sum_{n=1}^N u_{n,x} + 10^3 \sum_{n=1}^N \left\| \boldsymbol{\sigma}_n \right\|_1 + 10^3 \sum \left\| \xi_n \right\|_1.
\end{align}
The absolute values ($L_1$ norms) are computed via the introduction of auxiliary variables. The choices of $10^{-2}$ and $10^3$ for penalization were found empirically to provide reliable convergence. The entire optimization problem is thus
\begin{mini*}
    {}{\eqref{equation_scp_objective}\quad\text{(objective function $J$)}}{}{}
    \addConstraint{\eqref{equation_dynamics_linear}}{}{\quad\text{(linearized dynamics)}}
    \addConstraint{\eqref{equation_sail_angle_limits}}{}{\quad\text{(solar sail angle limits)}}
    \addConstraint{\eqref{equation_sail_transverse_norm}}{}{\quad\text{(solar sail transverse SOC norm)}}
    \addConstraint{\eqref{equation_sail_transverse_taylor}}{}{\quad\text{(solar sail transverse linearization)}}
    \addConstraint{\eqref{equation_state_trust_regions}}{}{\quad\text{(state hard trust regions)}}
    \addConstraint{\eqref{equation_eccentricity_constraint}}{}{\quad\text{(maximum eccentricity)}}
    \addConstraint{\eqref{equation_semi_major_axis_constraint}}{}{\quad\text{(semi-major axis bounds)}}
    \label{equation_full_scp_problem}.
\end{mini*}
The \gls{SCP} process repeatedly solves this convex problem and updates the linearized constraints \eqref{equation_dynamics_linear}, \eqref{equation_sail_transverse_taylor}, \eqref{equation_eccentricity_constraint} and \eqref{equation_semi_major_axis_constraint} with the optimal solution from the previous iteration. The convergence of the algorithm is determined by the accuracy of the linearization compared to the true dynamics obtained from numerical propagation. In terms of implementation, \texttt{JuMP.jl} \citep{lubinJuMPRecentImprovements2023} is used to create and modify the convex problems, and MOSEK \citep{mosekapsMOSEKOptimizerAPI2025} is used to solve them.

\section{Results and Discussion}
\label{sec_results_discussion}

The results from the application of the translation-based \gls{MISOCP} station-keeping algorithm to a solar sail spacecraft in a polar \gls{eLLO} are presented. A grid search is performed over a range of initial orbital configurations to identify the best options for station-keeping with a solar sail, and the \gls{SCP} formulation is used to produce high-fidelity assessments of the identified configurations. The main focus is to improve understanding of the controllability of solar sails in these orbits, the effects of the Sun--Moon phase angle, and to identify the best orbital configurations for station-keeping.

\subsection{Spacecraft and mission definition}
\label{sec_spacecraft}

This work considers a spacecraft equipped with a solar sail in a polar or near-polar \gls{eLLO} around the Moon. The parameters of the solar sail are based on those of NASA's NEA Scout mission \citep{lantoineTrajectoryManeuverDesign2024}, which flew as a secondary payload on Artemis I, although communication with the spacecraft was not established after launch.

\begin{table}[h]
\centering
\caption{Spacecraft and solar sail parameters.}
\label{table_spacecraft_parameters}
\begin{tabular}{l l}
\toprule
Parameter & Value \\
\midrule
Spacecraft mass $m$ & 11.629~kg \\
Sail area $A$ & 84.6~m$^2$ \\
Specular reflection coefficient $\mu$ & 0.40495 \\
Diffuse reflection coefficient $\nu$ & 0.014957 \\
Solar flux $C$ at 1~AU & $4.5391 \cdot 10^{-6}$~N/m$^2$ \\
Characteristic acceleration at 1~AU & 0.0607~mm/s$^2$ \\
Sail cone angle limits & [0$^\circ$, 75$^\circ$] \\
\bottomrule
\end{tabular}
\end{table}

Table~\ref{table_spacecraft_parameters} summarizes the key spacecraft and solar sail parameters used in this work. The sail is a non-ideal flat-plate with both specular and diffuse reflection components, and has an area-to-mass ratio of approximately 7.27~m$^2$/kg. The characteristic acceleration $a_c$ is defined as the maximum acceleration that the solar sail can achieve at 1~AU when the sail normal is aligned with the sunlight direction. In contrast to other studies on the NEA Scout spacecraft \citep{lantoineTrajectoryManeuverDesign2024, oguriLosslessControlConvexFormulation2024}, the sail cone angle limits have been increased. This enhances the spacecraft's controllability, particularly in orbits that are directly aligned with the Sun--Moon line.

\begin{table}[h]
\centering
\caption{Nominal parameters for the \gls{LRO}-inspired case study.}
\label{table_orbital_parameters}
\begin{tabular}{l l}
\toprule
Parameter & Value \\
\midrule
Semi-major axis & 1787.4~km \\
Eccentricity & 0.0 \\
Inclination & 92.5$^\circ$ \\
Longitude of ascending node & 0$^\circ$ \\
Argument of periapsis & 90$^\circ$ \\
True anomaly & 0$^\circ$ \\
Mean altitude & 50~km \\
Station-keeping periapsis bounds & [25, 50]~km \\
Station-keeping eccentricity bounds & [0, 0.01399] \\
\bottomrule
\end{tabular}
\end{table}

An \gls{LRO}-inspired polar orbit is selected as the nominal case study for this work. Table~\ref{table_orbital_parameters} summarizes the orbit and station-keeping parameters used in this work. A start time of January 1, 2026, at 00:00 UTC is used for all simulations unless otherwise stated, which defines the Sun--Moon phase angle and orientation of the lunar surface. The eccentricity bounds for the intended station-keeping region are computed based on an orbit with a periapsis altitude of 25~km and a mean altitude of 50~km. This is slightly less than the 30~km periapsis altitude bounds used for \gls{LRO} analysis \citep{beckmanStationkeepingLunarReconnaissance2007} but will still provide a significant margin against lunar impact.

\subsection{Controllability analysis of solar sail in eLLOs}
\label{sec_controllability_analysis}

The controllability of the solar sail in the \gls{eLLO} environment is analyzed to assess the effects of shadowing on the spacecraft trajectory, as well as the influence of the orbit--Sun phase angle. The reference sail and orbit parameters from Table~\ref{table_spacecraft_parameters} and Table~\ref{table_orbital_parameters} are used. In general, the results of this analysis are expected to map to other configurations of orbital parameters and solar sails, but will remain strongly geometry dependent.

\begin{figure}[h!]
\centering
\includegraphics[width=\textwidth]{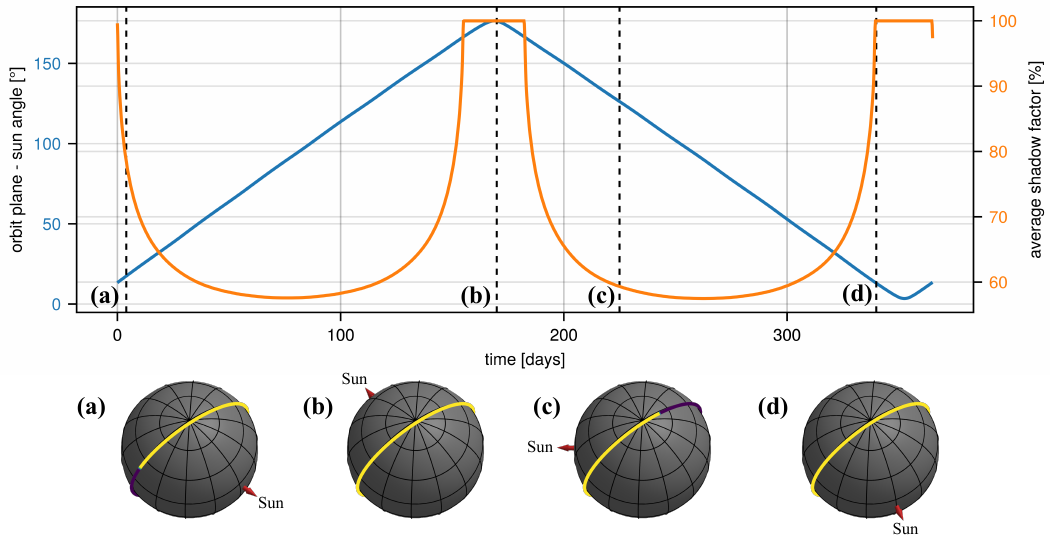}
\caption{Average shadow factor over one orbital revolution and corresponding orbit--Sun phase angle for start epochs spanning 1~year; panels (a)--(d) highlight representative Sun--Moon--orbit geometries.}
\label{figure_sail_shadowing_phase}
\end{figure}

A study on the duration of shadowing is then conducted, the results of which are illustrated in Fig.~\ref{figure_sail_shadowing_phase}. Over a range of starting times spanning 1 year from January 1, 2026, at 00:00 UTC, the initial condition is propagated for one orbital revolution. The average shadow factor corresponds to the fraction of time spent in sunlight, while the orbit--Sun phase angle corresponds to the angle between the orbital angular momentum vector and the Moon--Sun line. Four examples (a--d) are also expanded upon, showing the geometric configuration of the orbit with respect to the Moon and Sun.

If the orbit--Sun angle is near 0$^\circ$ or 180$^\circ$, the spacecraft experiences continuous sunlight, while near 90$^\circ$ or 270$^\circ$ the spacecraft experiences the maximum shadowing. Because of this, unless the longitude of the ascending node can be swept rapidly (e.g., as in Sun-synchronous orbits), there will always be extended periods of time where the solar sail experiences significant shadowing throughout the orbit. Shadowing is of particular concern to solar sails, as not only is the acceleration magnitude reduced, but the points in the orbit where the sail can produce acceleration are limited in an uneven manner. This can quickly induce undesirable changes to the other orbital elements, not just the eccentricity vector.

\begin{figure}[h!]
\centering
\includegraphics[width=\textwidth]{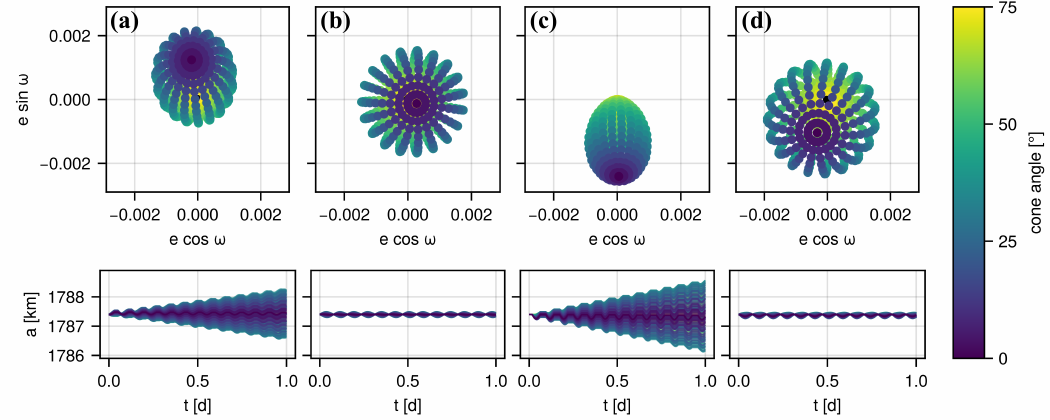}
\caption{One-day reachability of the eccentricity vector and semi-major axis under central-body-plus-\gls{SRP} dynamics for the representative shadowing cases in Fig.~\ref{figure_sail_shadowing_phase}.}
\label{figure_sail_reachability}
\end{figure}

For the cases (a--d) in Fig.~\ref{figure_sail_shadowing_phase}, a reachability analysis is performed to understand how the different shadowing conditions can affect the controllability of the solar sail. Taking the start epoch of each case, the initial condition is propagated for one day in the presence of the central two-body term and the solar sail \gls{SRP} only in order to isolate the effect of the solar sail. For this analysis, the solar sail cone and clock angles are discretized into 20 increments each, yielding a total of 400 control configurations.

The results are illustrated in Fig.~\ref{figure_sail_reachability}. Importantly, for all cases, the solar sail is almost able to keep the eccentricity vector steady at the origin, indicating that without other perturbations, the solar sail is capable of station-keeping itself. However, this holds only when larger cone angles are permitted. In particular, case (c), which experiences the most shadowing, begins to struggle to maintain control even with the larger cone angles, exhibiting a continuous downward drift in the eccentricity vector. Since conditions similar to case (c) are expected to occur for large portions of the year, this indicates that larger cone angles may be required for effective station-keeping close to the lunar surface. This was the primary motivation for increasing the maximum cone angle to 75$^\circ$ in comparison with previous studies with the NEA Scout solar sail design \citep{lantoineTrajectoryManeuverDesign2024, oguriLosslessControlConvexFormulation2024}, with a maximum cone angle of 50$^\circ$.

The other important orbital element to consider is the semi-major axis. The \gls{MISOCP} algorithm only considers the control of the eccentricity vector, but changes in the semi-major axis can be independent of changes in the eccentricity vector, so it is important to consider its controllability. As may be expected by the orbital geometry, both cases (b) and (d) are able to keep the semi-major axis steady as they do not exhibit shadowing. Cases (a) and (c), however, both exhibit large variations in semi-major axis, especially when taking into account the short timescale of one day. However, it appears that this variation may be controlled to some extent by the clock angle, since the range of cone angles appears relatively symmetric about the drifts in semi-major axis. Therefore, this analysis indicates that although semi-major axis control may be challenging, it appears to be feasible to control with a solar sail in these orbits.

This analysis provides important insight into the limitations imposed by the solar sail area-to-mass ratio on stable station-keeping strategies. Increasing the area-to-mass ratio will effectively magnify the spacecraft eccentricity vector reachability and therefore enhance the station-keeping performance during controllable periods. However, during less-controllable intervals, even larger cone angles will be required to reduce drift. Consequently, without either an increase in the maximum allowable cone angle or a mechanism to modulate the sail area, spacecraft designs with area-to-mass ratios exceeding that of NEA Scout may begin to destabilize the station-keeping strategy.

\begin{figure}[h!]
\centering
\includegraphics[width=\textwidth]{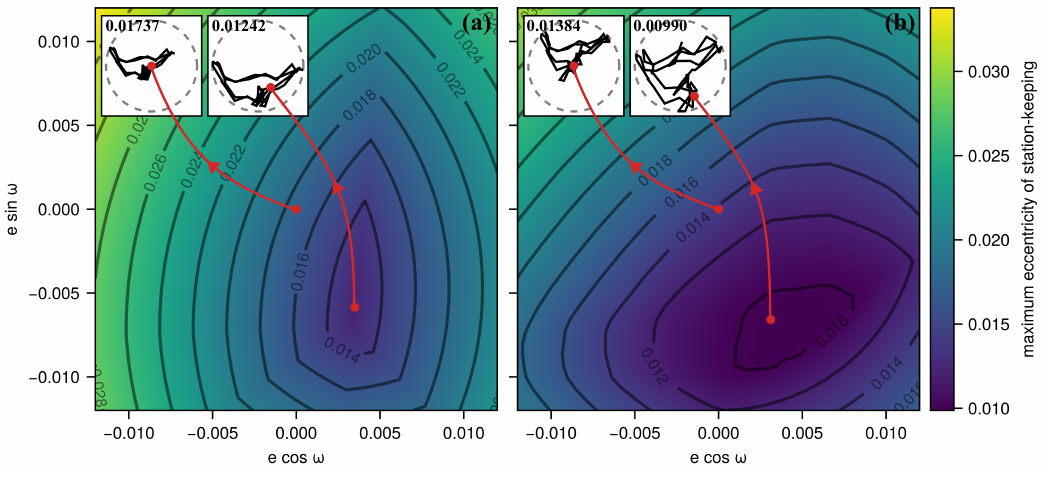}
\caption{Maximum eccentricity over 60 days as a function of the initial eccentricity-vector location for ballistic propagation and translation-based \gls{MISOCP} control with solar sailing.}
\label{figure_translation_minimum_radius}
\end{figure}

\begin{figure}[h!]
\centering
\includegraphics[width=\textwidth]{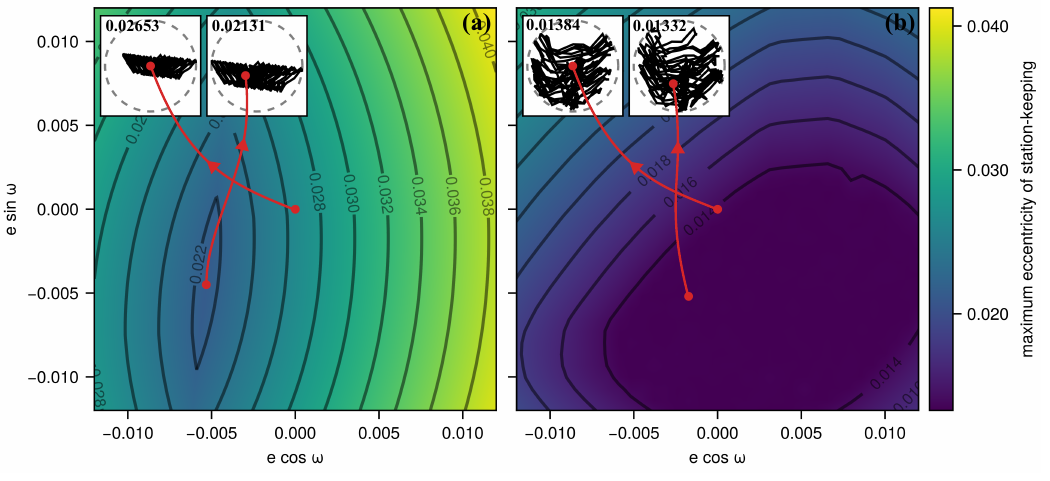}
\caption{Maximum eccentricity over 1~year as a function of the initial eccentricity-vector location for ballistic propagation and translation-based \gls{MISOCP} control with solar sailing.}
\label{figure_translation_minimum_radius_1year}
\end{figure}

Finally, to evaluate the performance of the proposed solar sail station-keeping strategy, a comparison with ballistic propagation without solar sailing is conducted. Fig.~\ref{figure_translation_minimum_radius} illustrates how the maximum eccentricity of the station-keeping trajectory changes depending on the choice of initial condition, over a 60-day propagation of the eccentricity vector. The left plot (a) shows the results for a propagation without the solar sail. Insets show the propagation of specific initial conditions and their maximum eccentricity values. These insets are generated for the reference case (the origin) and for the best-found initial condition. There are regions of the eccentricity-vector plane that are superior to others for station-keeping, with the best-found initial condition resulting in an approximate 28\% reduction in maximum eccentricity compared to the reference propagation.

In contrast, the right plot (b) shows the results when solar sailing is included. For each starting location in this plot, the \gls{MISOCP} formulation is solved to find the optimal solar sail control strategy that minimizes the maximum eccentricity. The insets again show the reference case and the best-found initial condition with solar sailing, but now each inset displays differences in the solar sail control profile, evident in the paths of the eccentricity vector. From this comparison, the inclusion of solar sailing significantly improves the station-keeping performance and allows for a much larger region of the eccentricity-vector plane to be effective for station-keeping. In this case, the best-found initial condition with solar sailing results in an approximate 43\% reduction in maximum eccentricity compared to the ballistic reference case without solar sailing.

For longer timescales, the benefits of solar sailing are even more pronounced, as drifts in the eccentricity vector can be more effectively countered. This is illustrated in Fig.~\ref{figure_translation_minimum_radius_1year}, which shows the same comparison as Fig.~\ref{figure_translation_minimum_radius} but instead for a 1-year propagation of the eccentricity vector. The ballistic propagation without solar sailing results in a significant drift in the eccentricity vector, preventing effective station-keeping even with the best choice of initial conditions. In contrast, the inclusion of solar sailing again significantly improves the station-keeping performance, with the best-found initial condition resulting in an approximate 50\% reduction in maximum eccentricity compared to the ballistic reference case without solar sailing. However, across all cases, the maximum eccentricity values are generally larger over longer timescales. This confirms the previous findings that for some orbital geometries, the solar sail control authority is limited due to shadowing effects or the orbit--Sun phase angle. This can have detrimental effects on the compactness of the eccentricity vector control over long durations. Nevertheless, with a 1-year timescale demonstrated, it is unlikely that further degradation to the maximum eccentricity would occur beyond this point, suggesting that long-duration station-keeping beyond 1 year may be feasible for favorable orbital configurations.

\subsection{Grid search over eLLO initial conditions}
\label{sec_grid_search}

\begin{figure}[t!]
\centering
\includegraphics[width=\textwidth]{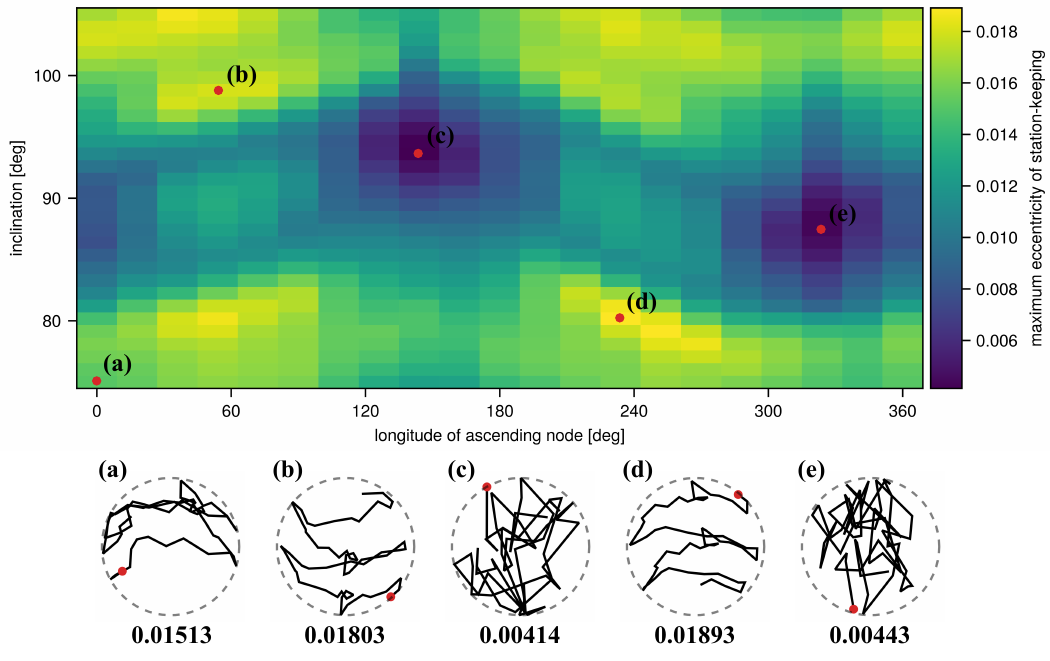}
\caption{Grid-search results for the 60-day case: minimum achievable maximum eccentricity versus inclination and longitude of the ascending node, with representative eccentricity-vector trajectories.}
\label{figure_grid_search_results_60days}
\end{figure}

A grid search is performed to find the optimal initial conditions over a range of initial orbital configurations. This is performed over the longitude of the ascending node $\Omega$ and the inclination $i$, while keeping the nominal semi-major axis from Table~\ref{table_orbital_parameters} fixed. The longitude of the ascending node is varied from 0$^\circ$ to 360$^\circ$ with a total of 20 increments, while the inclination is varied from 75$^\circ$ to 105$^\circ$ with a total of 30 increments. This results in a total of 600 initial orbital configurations to test. For each of these configurations, the \gls{MISOCP} formulation is solved to find the optimal solar sail controls and initial eccentricity vector location that minimizes the maximum eccentricity. The grid search is performed for two different station-keeping durations: a shorter 60-day case and a longer 1-year case.

For the 60-day case, the results are illustrated in Fig.~\ref{figure_grid_search_results_60days}. The top panel shows a heatmap of the (minimum) maximum eccentricity for each initial orbital configuration. Strong dependencies on both the inclination and the longitude of the ascending node are evident. In particular, there appear to be two weak horizontal bands, with inclinations near 85$^\circ$ and 95$^\circ$, where there seems to be slightly better station-keeping performance. These bands also occur in previous analyses of \glspl{eLLO} for impulsive station-keeping strategies \citep{yarndleyOriginsApplicationTranslation2026}. However, even more pronounced are the vertical bands, which appear at approximately 145$^\circ$ and 325$^\circ$ in longitude of the ascending node. These bands appear to correspond to configurations where the orbit plane and Sun are nearly aligned (see Fig.~\ref{figure_sail_shadowing_phase}), indicating that the orbit--Sun phase angle is a critical factor in determining the controllability and, correspondingly, the station-keeping performance of the solar sail in these orbits.

The bottom panels (a--e) show the resulting eccentricity vector trajectories for selected initial configurations. Among these panels, panels (b) and (d) correspond to some of the worst-performing configurations and clearly exhibit a large uncontrolled drift in the eccentricity vector. In contrast, panels (c) and (e) correspond to some of the best-performing configurations and exhibit tight control of the eccentricity vector around the origin. In fact, panel (c) corresponds to the best-found configuration, with an inclination of 93.62$^\circ$ and a longitude of the ascending node of 144.0$^\circ$, resulting in a maximum eccentricity of only 0.00414 over the 60-day duration.

\begin{figure}[t!]
\centering
\includegraphics[width=\textwidth]{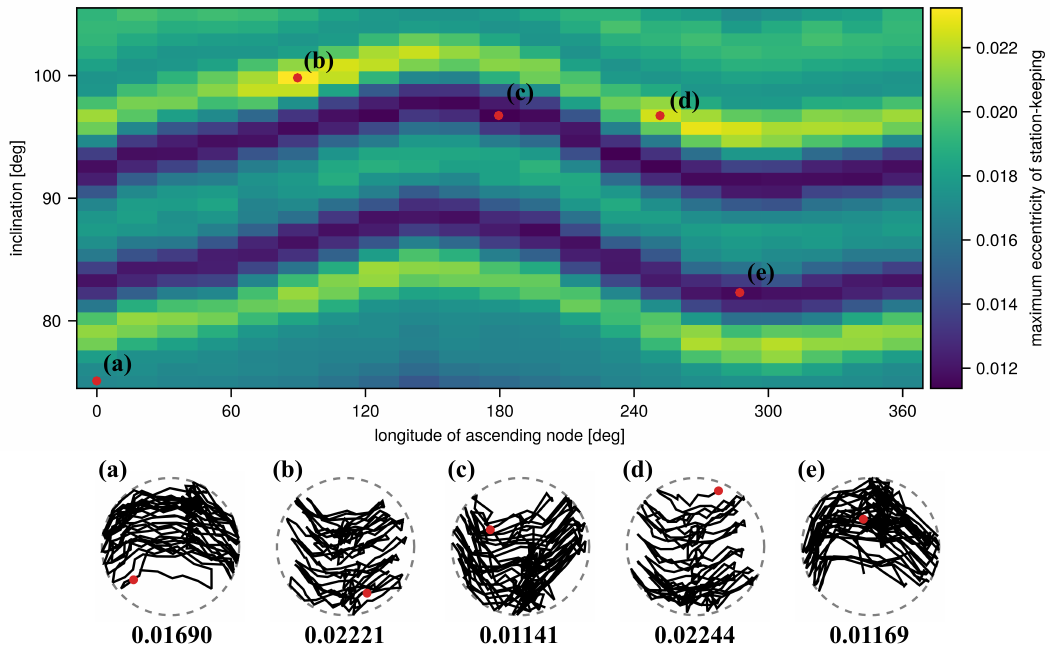}
\caption{Grid-search results for the 1-year case: minimum achievable maximum eccentricity versus inclination and longitude of the ascending node, with representative eccentricity-vector trajectories.}
\label{figure_grid_search_results_1year}
\end{figure}

For the longer 1-year case, the results are illustrated in Fig.~\ref{figure_grid_search_results_1year}, with the same format as Fig.~\ref{figure_grid_search_results_60days}. Only some of the trends in the shorter 60-day case remain. The horizontal bands are now much more pronounced and have a clear dependence on both inclination and the longitude of the ascending node. The vertical bands are no longer visible, indicating that the orbit--Sun phase angle is less critical over longer durations. This is likely because, over a full year, the orbit will experience all possible Sun--Moon geometries, so the initial geometry, particularly the orbit--Sun phase angle (which is defined by the longitude of the ascending node), becomes less important.

In contrast to the 60-day case, the maximum eccentricities are larger due to the longer duration. This is particularly evident in the best cases, where they now exhibit periods of poor controllability due to shadowing effects or unfavorable orbit--Sun phase angles. Panels (b) and (d) correspond to some of the worst-performing configurations, which exhibit large uncontrolled drifts in the eccentricity vector. Panels (c) and (e) demonstrate some of the best-performing configurations and exhibit tight control of the eccentricity vector around the origin. Panel (c) corresponds to the best-found configuration, with an inclination of 96.72$^\circ$ and a longitude of the ascending node of 180.0$^\circ$, resulting in a maximum eccentricity of 0.01141 over the 1-year duration. This satisfies the station-keeping eccentricity bounds from Table~\ref{table_orbital_parameters} and assuming a fixed semi-major axis would correspond to a minimum periapsis of 30.4~km.

\subsection{Refinement of best orbital configurations}
\label{sec_best_orbits}

As the final step, the best-found orbital configurations from the grid search in Sec.~\ref{sec_grid_search} are refined using the high-fidelity \gls{SCP} formulation. This refinement verifies that the trajectories are truly feasible, since the \gls{MISOCP} formulation relies on the imperfect translation theorem approximation. It also ensures that there is no significant drift in the semi-major axis, for which a band of $\pm$5~km around the nominal semi-major axis is enforced in the \gls{SCP}.

For both tested cases, the \gls{SCP} process is warm-started using the results from the \gls{MISOCP} formulation, with node and control updates occurring once per day. At each node the eccentricity vector is initialized at the locations found from the \gls{MISOCP} and the semi-major axis is reset to the nominal value. The corresponding virtual controls for the warm-start are then computed. Convergence of the \gls{SCP} algorithm occurs within 30 iterations for both cases.

\begin{figure}[t!]
\centering
\includegraphics[width=\textwidth]{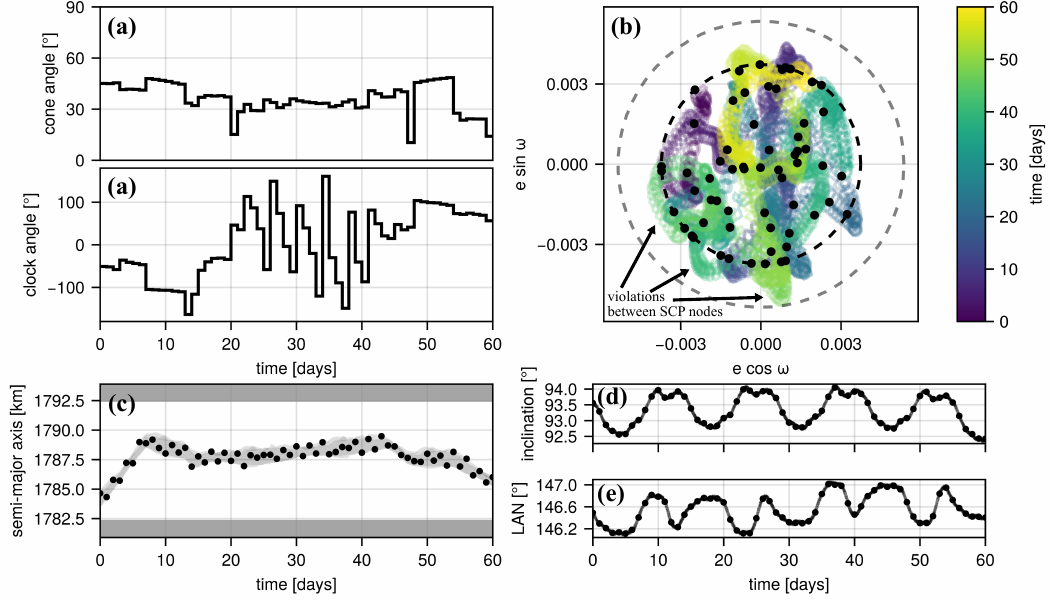}
\caption{\gls{SCP}-refined 60-day station-keeping solution for the best grid-search configuration, including sail attitude, eccentricity-vector evolution, and orbital-element histories.}
\label{figure_results_scp_60days}
\end{figure}

Fig.~\ref{figure_results_scp_60days} presents the results of the \gls{SCP} for the 60-day case, which has an average cone angle of 36.12$^\circ$ and completes a total of approximately 750 orbital revolutions. In all relevant panels, the line color corresponds to time. The right panel (b) shows the evolution of the eccentricity vector over time, demonstrating the tight control achieved around the origin. The nodes of the \gls{SCP} are shown as black dots and the resulting maximum eccentricity is shown as a dark dashed circle which encompasses all of the nodes. Of the nodes, the maximum eccentricity is 0.00373, which is slightly better than the \gls{MISOCP} result of 0.00414. This improvement is likely due to a combination of factors, including the use of the true dynamics rather than a translation approximation and the finer control resolution enabled by the \gls{SCP} formulation. Additionally, there is an apparent difference between the maximum eccentricity encountered (shown as a gray dashed circle) and the maximum eccentricity at the nodes. When considering these node-to-node violations, the maximum eccentricity increases to 0.00535. This discrepancy occurs because the maximum eccentricity constraint is only checked and enforced at the discrete nodes in both the \gls{MISOCP} and \gls{SCP}. Nevertheless, for all metrics, the maximum eccentricity remains well within the station-keeping bounds.

The control profile for the solar sail is shown in the left panels marked (a). Over the duration of station-keeping, the sail cone angle varies between approximately 0$^\circ$ and 50$^\circ$; in this case, the sail cone angle limit is not binding. The cone angle generally varies relatively smoothly over time, with more rapid changes occurring in the clock angle. This `fishtailing' behavior is typical of solar sail control profiles \citep{lantoineTrajectoryManeuverDesign2024}, and results from the underactuated nature of the solar sail, where rapid changes in clock angle are often required to maintain control authority. This could be mitigated to some extent by introducing penalties on the cone and clock angle rates, which can be relatively easily implemented within \gls{SCP} but are not considered in this work.

The bottom panel (c) shows the variation in the semi-major axis over time. The semi-major axis remains well within the prescribed bounds of $\pm$5~km around the nominal value of 1787.4~km, and the constraint is not binding at any node. Therefore, in this case, the semi-major axis constraint is not critical to the station-keeping performance. The inclination, shown in panel (d), remains relatively near the nominal grid-search value of 93.62$^\circ$, but due to the free initial condition of the \gls{SCP} it is no longer the exact starting value. Similarly, the initial longitude of the ascending node, shown in panel (e), is not exactly 144.0$^\circ$, but it remains close to this value with a small increasing drift over time. The grid search, therefore, provides initial conditions that are sufficiently close for the \gls{SCP} process to refine into a high-quality station-keeping solution.

\begin{figure}[t!]
\centering
\includegraphics[width=\textwidth]{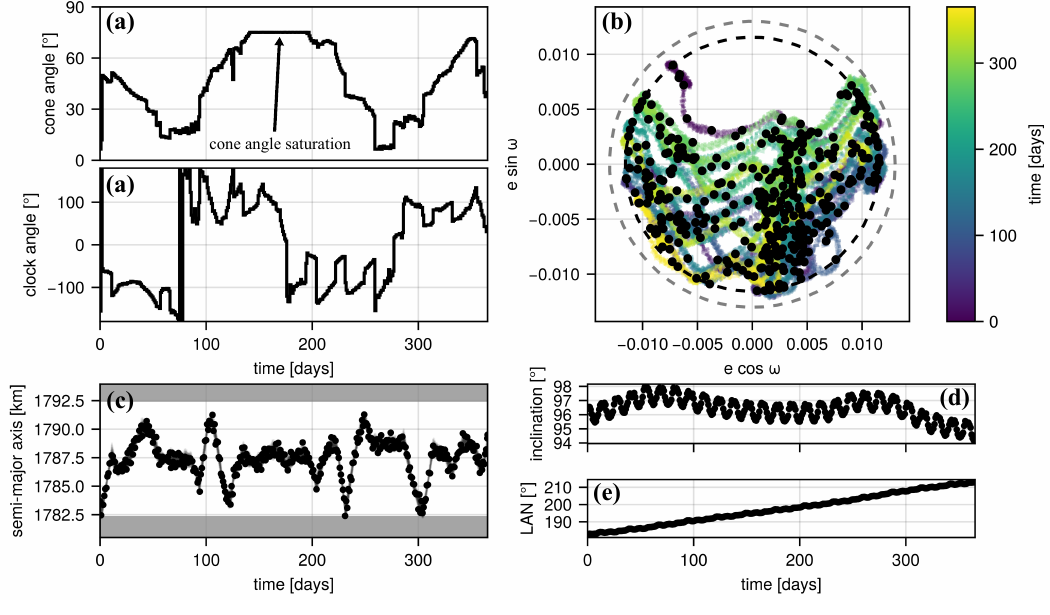}
\caption{\gls{SCP}-refined 1-year station-keeping solution for the best grid-search configuration, including sail attitude, eccentricity-vector evolution, and orbital-element histories.}
\label{figure_results_scp_1year}
\end{figure}

For the 1-year case, Fig.~\ref{figure_results_scp_1year} presents the results in the same format as Fig.~\ref{figure_results_scp_60days}. The trajectory completes approximately 4650 orbital revolutions and has an average cone angle of 46.30$^\circ$. In this case, the maximum eccentricity at the nodes is 0.01157, which is very slightly worse than the \gls{MISOCP} result of 0.01141. If the violations between the nodes are also considered, the maximum eccentricity increases to 0.01298. Both of these values represent performance degradations compared to the \gls{MISOCP} results. This degradation is not unexpected; it is likely due to the increased challenge of maintaining control over the longer duration, errors in the translation approximation, and the binding of the semi-major axis constraints, which is completely ignored within the \gls{MISOCP}. Nevertheless, the maximum eccentricity remains within the station-keeping bounds from Table~\ref{table_orbital_parameters}, corresponding to a minimum periapsis of 26.8~km assuming the nominal semi-major axis.

Compared to the 60-day case, the cone angle in the 1-year case is generally larger throughout station-keeping, and there is a significant period during which it saturates at 75$^\circ$. This occurs during a period where the orbit--Sun phase angle is favorable to station-keeping (see Fig.~\ref{figure_sail_shadowing_phase}), which is consistent with a period of low mean-motion drift of the eccentricity vector due to the natural dynamics. In this region, the solar sail would only be required to make small corrections, resulting in a larger cone angle, so a small cone angle could then be detrimental to the station-keeping performance.

In this case, the semi-major axis constraint is much more critical to the station-keeping performance. The semi-major axis drifts significantly over the course of the year, and there are several periods during which the constraint is binding at the nodes, forcing them to the lower bound of 1782.4~km. Without the constraint, the semi-major axis would drift significantly, potentially leading to minimum altitude violations. The longer timescale of the 1-year case also allows for more significant drifts in the inclination and longitude of the ascending node, which are not constrained in the \gls{SCP}. The inclination drifts significantly more than in the 60-day case, but remains near the nominal value of 96.72$^\circ$. However, the longitude of the ascending node drifts significantly but in a constant increasing manner.

\subsection{Impact of control update frequency}
\label{sec_control_update_frequency}

A once-per-day update frequency for the solar sail pointing angle may not be operationally feasible, depending on the mission design and communication capabilities. Therefore, it is important to understand how station-keeping performance degrades as the control update frequency decreases. To assess the sensitivity of the refined station-keeping strategy to the frequency of the control updates, the 1-year case from Sec.~\ref{sec_best_orbits} is re-optimized with a range of different control update intervals.

This is achieved using the same guess control history (from the \gls{MISOCP}) but with resampling to match the new update intervals. The resampling is performed by taking the instantaneous control from the baseline solution at each new update time. For cases where the update interval does not align evenly with the 1-year duration of the baseline solution, a slightly shorter segment is used at the end to ensure that the final time still corresponds to exactly one year. The \gls{SCP} is slightly modified from the formulation presented in Sec.~\ref{sec_scp_translation} to allow the state and control nodes to be decoupled, so the control update intervals can be independent of the state node intervals. In this way, the state nodes can remain at a 1-day cadence while the control nodes are resampled to the new update intervals, which ensures a fair performance comparison across the different update intervals.

This control then informs the warm-start of the \gls{SCP} process, which is then re-optimized to find the best control profile for the new update interval that minimizes the maximum eccentricity. This process is repeated for the update intervals of 0.5, 1, 4, 7, 10, and 28 days, with the 1-day case exactly corresponding to the baseline solution from Sec.~\ref{sec_best_orbits}.

\begin{figure}[t!]
\centering
\includegraphics[width=\textwidth]{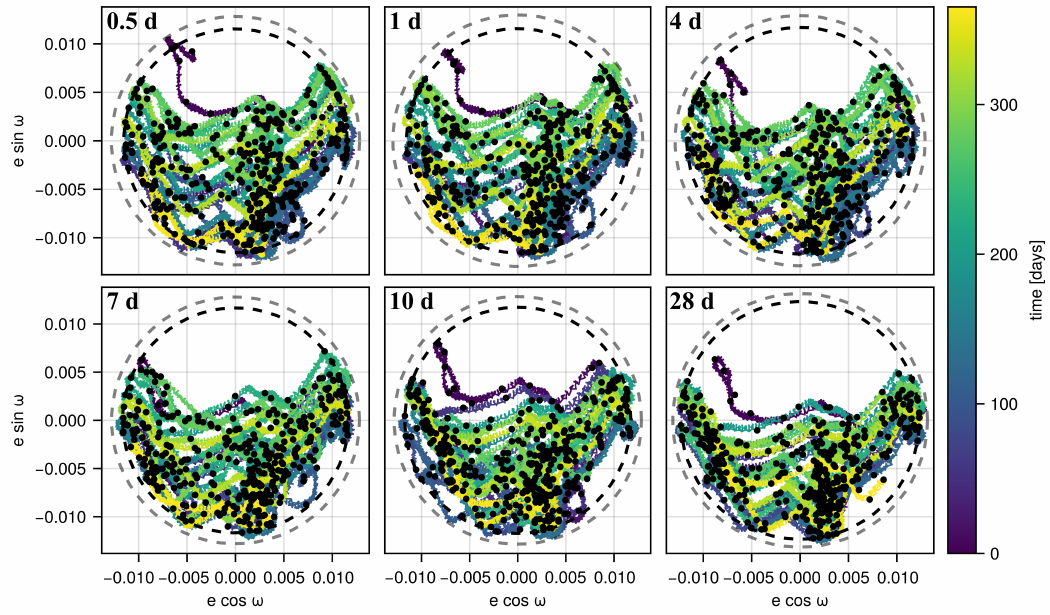}
\caption{\gls{SCP}-refined eccentricity-vector trajectories for control update intervals from 0.5 to 28 days; black dots denote the state nodes.}
\label{figure_control_update_frequency}
\end{figure}

\begin{table}[t!]
\centering
\caption{Impact of the control update frequency on station-keeping performance.}
\label{table_control_update_frequency}
\begin{tabular}{lrrrrr}
\toprule
Control update & Max $e$ & Max $e$ & Min $a$ & Max $a$ & Average cone \\
interval & (nodes) & (all) & [km] & [km] & angle [$^\circ$] \\
\midrule
0.5~d  & 0.01154 & 0.01282 & 1781.22 & 1792.14 & 45.74 \\
1~d    & 0.01157 & 0.01298 & 1782.05 & 1791.62 & 46.30 \\
4~d    & 0.01169 & 0.01291 & 1781.67 & 1792.45 & 48.01 \\
7~d    & 0.01167 & 0.01280 & 1781.90 & 1792.36 & 50.32 \\
10~d   & 0.01174 & 0.01285 & 1781.81 & 1792.39 & 52.44 \\
28~d   & 0.01233 & 0.01315 & 1782.30 & 1791.18 & 64.68 \\
\bottomrule
\end{tabular}
\end{table}

The resulting eccentricity-vector trajectories across all control update interval cases are illustrated in Fig.~\ref{figure_control_update_frequency}. The state nodes of the \gls{SCP} are shown as black dots, which have a frequency of 1 day. The maximum eccentricity at the nodes is shown as a dark dashed circle while the maximum eccentricity across the entire trajectory is shown as a light dashed circle. Table~\ref{table_control_update_frequency} summarizes the key station-keeping metrics across the different control update intervals.

These results clearly demonstrate that the station-keeping performance is only weakly sensitive to the control update interval over a broad range. Across the cases from 0.5 to 10 days, the eccentricity-vector evolution remains very similar to the baseline 1-day solution, with only small differences in the associated semi-major axis and eccentricity excursions. For the nominal 1-year case, this suggests that the dominant control requirements vary on a timescale longer than the 1-day update cadence used in the \gls{SCP}.

A noticeable degradation in the maximum eccentricity begins to appear with the 28-day case. The small increase indicates substantial operational flexibility in the choice of update interval. However, acceptable solutions were not obtained for update intervals beyond 28 days, including a 56-day case which was tested but did not converge, suggesting that although the control update frequency can be increased significantly, there is a fundamental limit to how far it can be increased without losing the ability to maintain station-keeping performance.

Across all cases, the semi-major axis constraint is binding in both the upper and lower bounds. In Table~\ref{table_control_update_frequency}, the semi-major axis values include the periods between the nodes, and the minimum and maximum values are only slightly violated. Interestingly, the average cone angle increases as the control update interval increases, which provides an important insight into the underlying control requirements, since the objective of the \gls{SCP} also includes terms that minimize the cone angle. As the control update interval increases, the solar sail must be angled more aggressively, which is likely due to the need to minimize detrimental drifts in the eccentricity vector induced by the solar sail. This is particularly evident in the 28-day case, where the average cone angle increases significantly to 64.68$^\circ$, which is close to the maximum cone angle limit of 75$^\circ$. This suggests that for update intervals longer than 28 days, the control requirements may become too aggressive for the solar sail to achieve, which could explain why acceptable solutions are not obtained for update intervals beyond 28 days.

\subsection{Sensitivity of station-keeping to uncertainties}
\label{sec_robustness_uncertainties}

The 60-day station-keeping case from Sec.~\ref{sec_best_orbits} is next used to assess the sensitivity of the solar-sail station-keeping strategy to navigation, modeling, and execution uncertainties. Both open- and closed-loop analyses are performed to assess the solar sail's ability to mitigate uncertainties in the station-keeping strategy.

Three sources of uncertainty are considered in these analyses: initial-state and navigation errors, slowly varying acceleration bias, and sail-normal pointing error. The initial-state and navigation error are modeled as zero-mean Gaussian perturbations in Cartesian coordinates, with a 1$\sigma$ position uncertainty of 250~m and a 1$\sigma$ velocity uncertainty of 5~mm/s, designed to be consistent with typical navigation errors for lunar missions \citep{mazaricoOrbitDeterminationLunar2018}. Navigation updates are taken with a cadence of 1-day, immediately prior to the 1-day control updates. The acceleration bias is modeled as a first-order Gauss--Markov process \citep{tapleyEstimationUnmodeledForces1975} and applied as a piecewise-constant perturbing acceleration over the 1-day control intervals. This has a correlation time of 5 days and a 1$\sigma$ magnitude of $1.0 \cdot 10^{-6}$~m/s$^2$, which is approximately 2\% of the nominal \gls{SRP} acceleration for the reference solar sail at 1~AU.

The sail normal pointing error is modeled as a zero-mean Gaussian rotation to the normal vector direction, with a 1$\sigma$ value of 1.0$^\circ$, after which the realized \gls{SRP} acceleration is evaluated using the perturbed normal vector. This pointing error is intended to capture a combination of attitude-control errors and execution errors in the solar sail deployment and control, and designed to be consistent with the pointing performance of the NEA Scout spacecraft \citep{lantoineTrajectoryManeuverDesign2024}. Selected values for all uncertainties are summarized in Table~\ref{table_uncertainty_parameters}.

\begin{table}[h]
\centering
\caption{Uncertainty values used in the closed-loop sensitivity analysis.}
\label{table_uncertainty_parameters}
\begin{tabular}{l l}
\toprule
Parameter & Value \\
\midrule
Position uncertainty (1$\sigma$) & 250~m \\
Velocity uncertainty (1$\sigma$) & 0.005~m/s \\
Acceleration drift correlation time & 5 days \\
Acceleration drift magnitude (1$\sigma$) & $1.0 \cdot 10^{-6}$~m/s$^2$ \\
Sail normal pointing error (1$\sigma$) & 1.0$^\circ$ \\
\bottomrule
\end{tabular}
\end{table}

For the open-loop case, the nominal control history from the \gls{SCP} solution is replayed under the perturbed conditions, without any feedback or correction. The closed-loop case employs a receding-horizon reference-tracking control strategy, in which a control profile is updated at regular intervals using the navigation estimate of the current state and the nominal \gls{SCP} solution.

The receding-horizon controller is implemented by repeatedly solving a short-horizon \gls{SCP} problem using the same dynamical model, lossless solar sail control constraints, and 1-day control cadence introduced in Sec.~\ref{sec_scp_translation}. The nominal 60-day \gls{SCP} solution serves as the reference trajectory. At each control update, a noisy estimate of the current Cartesian state is generated and imposed as the initial condition for a 1-day tracking problem. The horizon is also 1 day, so only local corrections are sought. The terminal target is taken from the reference solution 1 day ahead, and the optimization is formulated to track the first three modified equinoctial elements $(p,f,g)$, which principally determine the semi-major axis and the eccentricity vector. The $p$ term was empirically weighted twice as heavily as the $f$ and $g$ terms to ensure control of the semi-major axis. Only the first control over the horizon is then executed before a new navigation update is received, and the optimization is repeated the next day. Because each correction represents only a local deviation from the converged nominal solution, a single iteration of \gls{SCP} is sufficient at each update. Consequently, the receding-horizon \gls{SCP} controller effectively reduces to a convex optimization solver, while preserving the convergence characteristics of the full \gls{SCP} formulation.

For each realization, the same sampled initial condition, acceleration-bias history, and pointing-error sequence are applied to both the open-loop and receding-horizon cases. This ensures that any performance difference is attributable to the controller rather than to a different uncertainty draw. A total of 100 realizations are generated.

\begin{figure}[t!]
\centering
\includegraphics[width=\textwidth]{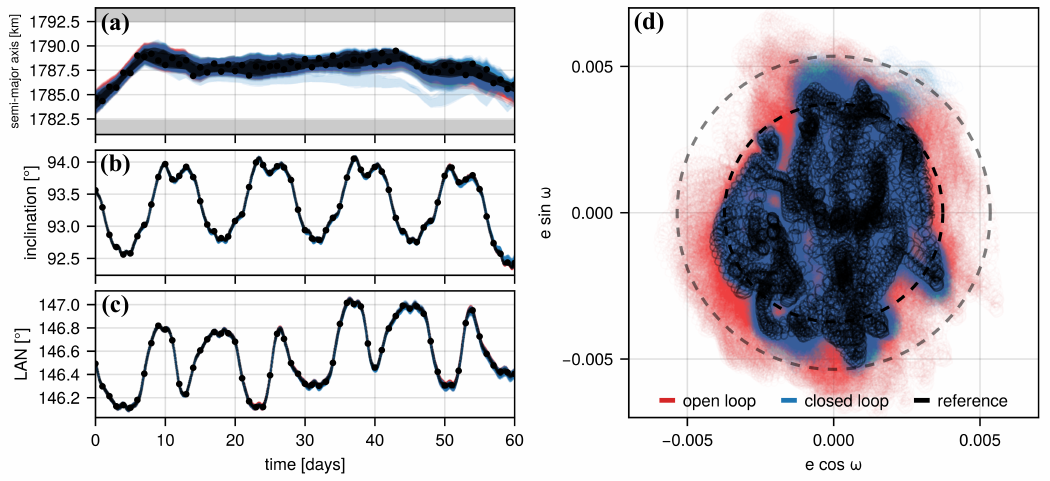}
\caption{Open- and closed-loop Monte Carlo trajectory ensembles for the 60-day case relative to the nominal reference trajectory.}
\label{figure_uncertainty_trajectories}
\end{figure}

\begin{figure}[t!]
\centering
\includegraphics[width=3.25in]{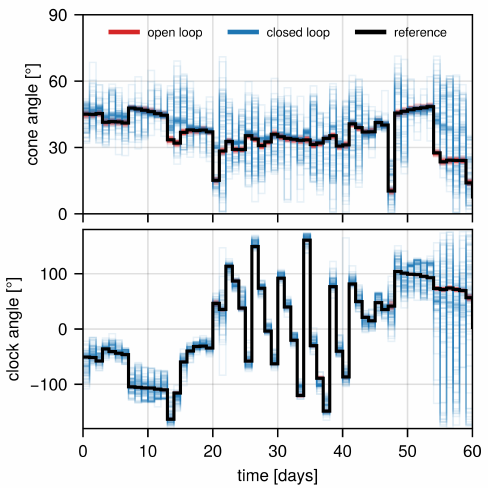}
\caption{Representative control profiles for open- and closed-loop Monte Carlo realizations for the 60-day case.}
\label{figure_uncertainty_controls}
\end{figure}

Fig.~\ref{figure_uncertainty_trajectories} illustrates the resulting trajectory ensembles for the open-loop and receding-horizon cases. The dominant effects of the uncertainties are evident in the semi-major axis and the eccentricity vector. The longitude of the ascending node and inclination are comparatively insensitive to the uncertainties. In particular, in the open-loop case, the eccentricity vector exhibits significant dispersion over the 60-day period, with many trajectories exceeding the station-keeping bounds. In contrast, the receding-horizon controller keeps the eccentricity vector more compact and close to the nominal solution. The semi-major axis exhibits comparatively little difference between the two cases. This behavior can be attributed to the high-frequency oscillations in the semi-major axis, which are difficult to regulate without targeting averaged orbital elements, and to the fact that open-loop trajectories already exhibit limited semi-major axis drift over the 60-day period.

In addition to the trajectory ensembles, Fig.~\ref{figure_uncertainty_controls} illustrates the control profiles for the open-loop and receding-horizon cases across the 100 realizations, compared with the nominal reference control history. The open-loop execution produces control histories that are slightly different from the reference profile, but are consistent with the 1.0$^\circ$ normal vector pointing error. In contrast, the receding-horizon controller generates a broader range of control responses that can deviate significantly from the nominal solution.

\begin{figure}[t!]
\centering
\includegraphics[width=\textwidth]{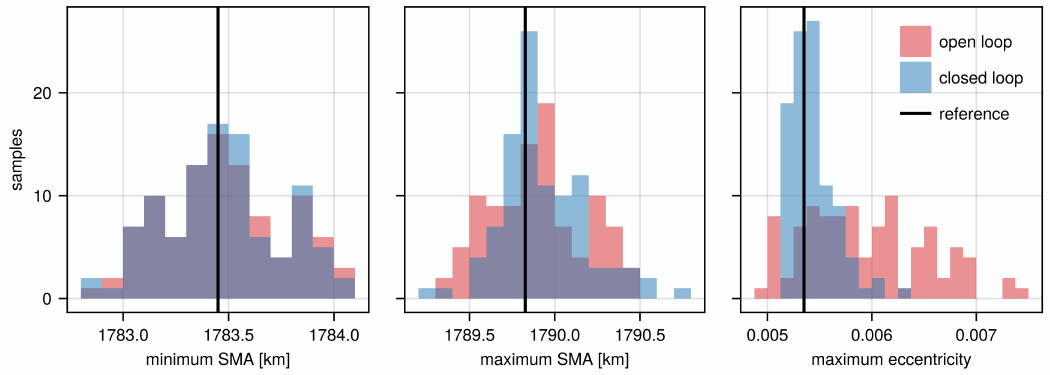}
\caption{Distributions of the minimum semi-major axis, maximum semi-major axis, and maximum eccentricity across the open- and closed-loop Monte Carlo realizations for the 60-day station-keeping case.}
\label{figure_uncertainty_histograms}
\end{figure}

As metrics to quantify the performance of the two approaches, the minimum and maximum semi-major axes and the maximum eccentricity are computed for each trajectory in the Monte Carlo ensemble. The resulting distributions are plotted in Fig.~\ref{figure_uncertainty_histograms}. Although the open-loop and receding-horizon distributions for the semi-major axis statistics are similar, the maximum eccentricity distribution is substantially improved by the receding-horizon controller, exhibiting both reduced dispersion and a large shift toward the reference solution.

Overall, the results indicate that the 60-day station-keeping solution is not highly sensitive to representative navigation, execution, and modeling uncertainties. More importantly, the incorporation of a daily receding-horizon controller substantially improves execution of the station-keeping strategy, demonstrating that the solar sail can not only maintain the desired orbit under nominal conditions but also mitigate the effects of disturbances and uncertainties encountered during execution. The low and well-controlled dispersion of the 60-day Monte Carlo ensemble further suggests that the methodology may remain effective over longer durations, such as within the 1-year station-keeping scenario.

\section{Conclusion}
\label{sec_conclusion}

This paper presents a station-keeping strategy for solar-sail spacecraft operating in \glspl{eLLO}. A key contribution is the extension of the translation theorem to include solar-sail dynamics, enabling the combined effects of the non-spherical lunar gravity field and solar sail \gls{SRP} to be represented as approximate translations of the eccentricity vector. This insight allows for the rapid evaluation of candidate station-keeping configurations through a \gls{MISOCP} formulation, which requires only a single high-fidelity reference propagation with the non-spherical lunar gravity field. The resulting translation-based solutions were then assessed using a \gls{SCP} framework warm-started using the \gls{MISOCP} results. The \gls{SCP} formulation incorporates a lossless convexification of the non-ideal solar-sail model, conical shadowing, high-order lunar spherical harmonics, and utilizes the \gls{MEE} elements to improve convergence with the highly multi-revolution station-keeping strategies.

The results demonstrate that even with the limited and highly underactuated control capability of a realistic solar sail, effective station-keeping is achievable in low-altitude polar \glspl{eLLO}. The controllability depends strongly on the Sun--Moon phase angle, cone angle limits, and initial orbital configuration. The successful station-keeping of an \gls{LRO}-inspired orbit with a spacecraft based on the NEA Scout mission is demonstrated for a 1-year duration, which suggests that even longer-duration station-keeping may be feasible.

Additional analysis indicates that station-keeping performance is only weakly sensitive to increasing the control update interval to up to approximately 28 days, providing a significant degree of operational flexibility. For the 60-day case, the strategy is not highly sensitive to the presence of state, acceleration, and pointing uncertainties. Overall, the proposed methodology improves understanding of solar-sail control in highly shadowed lunar orbits and provides an efficient design tool for future missions requiring long-duration observations near the lunar surface.

\section*{Acknowledgments}
Part of the research presented in this paper has been carried out at the Jet Propulsion Laboratory, California Institute of Technology, under a contract with the National Aeronautics and Space Administration. The contributions of authors Jack Yarndley and Roberto Armellin are partially supported by the Royal Society Te Ap\=arangi through the Catalyst: Seeding grant titled ``Advanced Cislunar Space Mission Design''.

\bibliographystyle{unsrtnat}
\bibliography{ref}

\end{document}